\documentclass[preprint2,tighten]{aastex6}
\pdfoutput=1 
\usepackage{amsmath,amstext}
\usepackage[T1]{fontenc}
\usepackage{apjfonts} 
\usepackage{gensymb}
\usepackage{verbatim}
\usepackage{graphicx,epstopdf}
\usepackage[figure,figure*]{hypcap}

\newcommand{\mb}{\,{\rm m_b}}
\newcommand{\mDM}{\,{\rm m_{DM}}}
\newcommand{\Mtot}{\,{\rm M_{tot}}}

\newcommand{\msun}{\,{\rm M_{\odot}}}

\newcommand{\Mstar}{M_*}
\newcommand{\Mstarcritrough}{10^{10.5}}
\newcommand{\Mdot}{{\dot M}}

\newcommand{\Rvir}{R_{\rm vir}}
\newcommand{\tcool}{t_{\rm cool}^{(s)}}

\newcommand{\tff}{t_{\rm ff}}

\newcommand{\Mhalo}{M_{\rm halo}}
\newcommand{\Mbh}{M_{\rm BH}}

\newcommand{\SigmaStar}{\Sigma^\star_{1 \rm kpc}}
\newcommand{\SigmaCrit}{\Sigma_{\rm crit}}

\newcommand{\vc}{v_{\rm c}}

\newcommand{\epsg}{\epsilon_{\rm gas}}

\newcommand{\tgrowth}{t_{\rm growth, BH}}
\newcommand{\tmodel}{t_{\rm threshold}}
\newcommand{\vesc}{v_{\rm esc}}
\newcommand{\tbursty}{t_{\rm bursty}}
\newcommand{\MdotBH}{\Mdot_{\rm BH}}

\newcommand{\thubble}{t_{\rm Hubble}}

\shorttitle{AASTeX6 Template}
\shortauthors{Byrne et al.}

\begin{document}

\title[Supermassive black hole growth in FIRE]{Stellar feedback-regulated black hole growth: driving factors from nuclear to halo scales}

\author{Lindsey Byrne\altaffilmark{1}$^*$}\email[*]{byrnelin@u.northwestern.edu}

\author{Claude-Andr{\'e} Faucher-Gigu{\`e}re\altaffilmark{1}}
\author{Jonathan Stern\altaffilmark{1,2}}
\author{Daniel Angl\'es-Alc\'azar\altaffilmark{3,4}}
\author{Sarah Wellons\altaffilmark{1,5}}
\author{Alexander B. Gurvich\altaffilmark{1}}
\author{Philip F. Hopkins \altaffilmark{6}}

\altaffiltext{1}{Department of Physics and Astronomy and CIERA, Northwestern University, Evanston, IL 60201, USA}
\altaffiltext{2}{School of Physics \& Astronomy, Tel Aviv University, Tel Aviv 69978, Israel}
\altaffiltext{3}{Department of Physics, University of Connecticut, 196 Auditorium Road, U-3046, Storrs, CT 06269-3046, USA}
\altaffiltext{4}{Center for Computational Astrophysics, Flatiron Institute, 162 5th Ave, New York, NY 10010, USA}
\altaffiltext{5}{Department of Astronomy, Van Vleck Observatory, Wesleyan University, 96 Foss Hill Drive, Middletown, CT 06459, USA}
\altaffiltext{6}{TAPIR, Mailcode 350-17, California Institute of Technology, Pasadena, CA 91125, USA}

\begin{abstract}
Several recent simulations of galaxy formation predict two main phases of supermassive black hole (BH) accretion: an early, highly intermittent phase (during which BHs are under-massive relative to local scaling relations), followed by a phase of accelerated growth.  We investigate physical factors that drive the transition in BH accretion in cosmological zoom-in simulations from the FIRE project, ranging from dwarf galaxies to galaxies sufficiently massive to host luminous quasars.  The simulations model multi-channel stellar feedback, but neglect AGN feedback. We show that multiple physical properties, including halo mass, galaxy stellar mass, and depth of the central gravitational potential correlate with accelerated BH fueling: constant thresholds in these properties are typically crossed within $\sim0.1$ Hubble time of accelerated BH fueling. Black hole masses increase sharply when the stellar surface density in the inner 1 kpc crosses a threshold $\SigmaStar \approx 10^{9.5} \msun {\rm kpc}^{-2}$, a characteristic value above which gravity prevents stellar feedback from ejecting gas, and similar to the value above which galaxies are observed to quench. We further show that accelerated BH growth correlates with the emergence of long-lived, thin gas discs, as well as with virialization of the inner circumgalactic medium. The halo mass $\Mhalo \sim 10^{12}$ M$_{\odot}$ and stellar mass $\Mstar \sim \Mstarcritrough$ M$_{\odot}$ at which BH growth accelerates correspond to $\sim L_{\star}$ galaxies. The fact that stellar feedback becomes inefficient at ejecting gas from the nucleus above this mass scale may play an important role in explaining why AGN feedback appears to be most important in galaxies above $L_{\star}$.
\end{abstract}


\section{Introduction}
\label{sec:intro}
Supermassive black holes (BHs) in galactic nuclei co-evolve with their host galaxies in ways which may significantly affect those galaxies, but which are not well understood. 
Indeed, active galactic nucleus (AGN) feedback is a core element of current galaxy formation theories, especially at the massive end \citep[e.g.,][]{SD2015_ARAA, NO2017_ARAA}. 
Observations have linked many properties of supermassive black holes to properties of their host galaxies, including kiloparsec-scale outflows of gas from galaxies hosting luminous quasars \citep[e.g.,][]{Feruglio2010QuasarOutflows,Rupke2013BREAKINGQSO,Cicone2014MassiveObservations,Fiore2017,Fluetsch2019} and scaling relations between black hole mass $\Mbh$ and the stellar properties of the host galaxy, such as galaxy mass, bulge mass, or velocity dispersion \citep[e.g.,][]{Magorrian1998TheCenters, Ferrarese2000AGalaxies, Tremaine2002, KH2013_ARAA, Sahu2019a}. 
Moreover, AGN feedback is the primary suspected driver of star formation quenching in massive galaxies, which is necessary to explain the blue/red color bi-modality \citep[e.g.,][]{Springel2005BlackGalaxies, Hopkins2008_redEs}. 
AGN feedback can operate through several different mechanisms, including kinetic winds, radiation, and powerful radio jets \citep[e.g.,][]{Fabian2012ObservationalFeedback}. 

In this paper, we use cosmological zoom-in simulations of galaxy formation from the FIRE project\footnote{See the FIRE project website at: http://fire.northwestern.edu.} \citep[][]{Hopkins2014GalaxiesFormation,Hopkins2018FIRE-2Formation} to investigate some of the physical processes that limit and/or drive the growth of massive black holes. 
The FIRE simulations provide a specific, high-resolution model to study the growth of massive black holes in the cosmological context, but our study is more broadly motivated by general trends that have been found in several recent simulations based on different codes and subgrid models. 
Specifically, a number of different simulations have found that SMBH growth is strongly inhibited at high redshift due to repeated gas ejection by stellar feedback \citep{Dubois2015BlackGrowth,Habouzit2017,Habouzit2021SupermassiveFunction,Bower2017TheEnd, Lapiner2021}. 
This results in nuclear black holes that, at low galaxy or bulge mass, can be order-of-magnitude under-massive relative to scaling relations measured at low redshift and primarily for more massive galaxies. 
It is only after some time, or after the host galaxy has grown sufficiently, that black holes ``catch up'' to masses expected from observed relationships in the local Universe.
\cite{Angles-Alcazar2017BlackNuclei} and \cite{Catmabacak2022BlackMergers} showed that a similar effect is seen in simulations of galaxies evolved with the FIRE model which, in detail, implements galaxy formation physics using quite different numerical models than the other simulations which found this same effect. 

The fact that delayed supermassive BH growth appears generic to very different numerical models suggests that it is due to fundamental processes that are common to most galaxy formation simulations. 
Thus, this appears to be a relatively robust prediction which may be realized in the real Universe. 
There is in fact some reported observational evidence for a ``break'' in the relationship between black hole mass and either the total galaxy stellar mass or the mass of the stellar spheroid at $\Mstar \sim 5 \times 10^{10}$ M$_{\odot}$ \citep[e.g.,][]{Graham2013THECANDIDATES, Reines2015RELATIONSUNIVERSE, Savorgnan2016, Sahu2019a}.
However, the presence of this break in the observations may depend on which galaxies are included in the analysis (e.g., early vs. late type galaxies) and exactly which galaxy property (e.g., stellar mass vs. stellar velocity dispersion $\sigma$) BH masses are compared to. 
As a result, some studies have reported no evidence of a break in scaling relations \citep[e.g.,][]{Schutte2019, Baldassare2020}, though these studies were based on relatively small data sets of dwarf galaxies. 

\begin{figure*}
\begin{center}
\includegraphics[width=\textwidth]{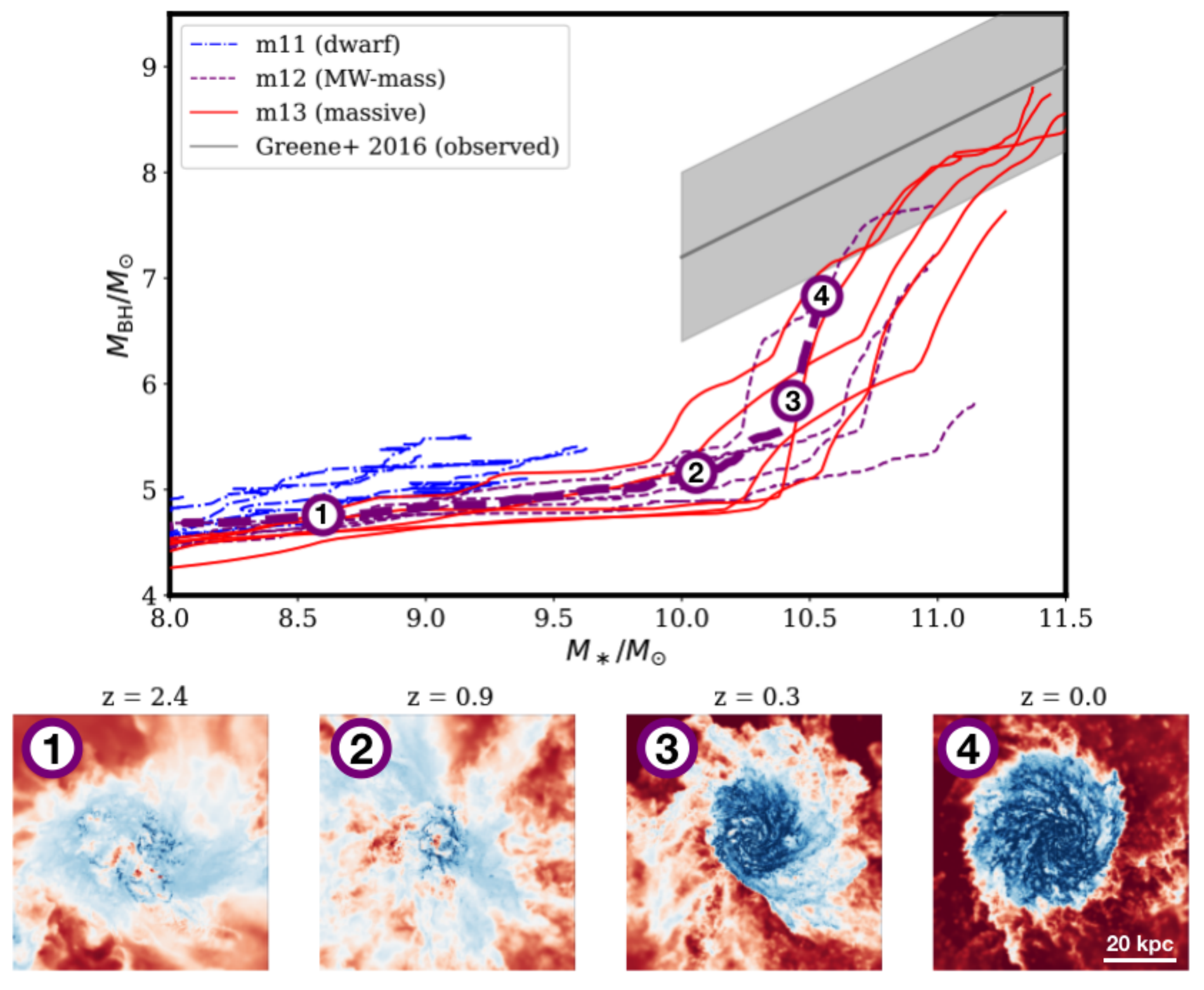}
\end{center}
\caption{The $\Mbh$-$\Mstar$ relation for all galaxies in our simulation sample, with gas temperature maps ($\log{T}$ is weighted by mass) shown at several snapshots for an m12 (Milky Way-mass) galaxy in the bottom row. Two phases of black hole growth create a kink in the relation between BH mass and the stellar mass of the galaxy at a mass scale of $\Mstar \sim \Mstarcritrough \msun$. 
Black holes accrete slowly at early times and then, in galaxies with sufficient stellar masses, BH fueling becomes more steady in a time-averaged sense. 
Note that the simulated BH masses found at the low-mass end are sensitive to the seed BH mass chosen for our analysis, but the overall kinked trajectories are robust.
The observational $\Mbh-\Mstar$ relation from \cite{Greene2016MEGAMASERGALAXIES} and its scatter are plotted in gray. The higher-mass simulated galaxies approach this relation at late times. }
\label{fig:MBH_Mbulge}
\end{figure*}

Depending on the stochasticity of BH fueling in the early phase, a break in the scaling relation could appear as increased scatter in scaling relations \citep[e.g.,][]{Lasker2016, Nguyen2019}. 
A recent study by \cite{Tillman_running_late} suggests that a break in the $\Mbh-\Mstar$ relation at a mass consistent with the FIRE simulations (see Fig. \ref{fig:MBH_Mbulge}) can provide a good fit to the observed quasar luminosity function, and may in fact be favored over models in which the scaling relation is purely linear. Ultimately, additional observations of low-mass galaxies will be required to resolve the issue of whether there are breaks in observed scaling relations. 
In this work, we focus on understanding the physics that drives the delayed SMBH growth predicted by simulations.

The potential implications of delayed SMBH growth are wide-ranging and include predictions for: (i) the redshift and mass evolution of black hole-galaxy scaling relations; (ii) the demographics of nuclear BHs in dwarf galaxies; (iii) AGN demographics; and (iv) the mergers of massive black hole that future gravitational wave experiments may detect \citep[][]{Bailes2021}. 

Furthermore, understanding the physical factors that eventually enable nuclear black holes to grow at an accelerated pace may shed light on the galaxy mass scale at which AGN feedback becomes most important.  
For example, \cite{Bower2017TheEnd} analysed simulations from the EAGLE project \citep[][]{Schaye2015} and found that accelerated BH growth occurs at roughly constant dark matter halo mass $M_{\rm h}\sim 10^{12}$ M$_{\odot}$, corresponding to the mass scale above which haloes become filled with hot, virialized gas \citep[the usual transition mass between ``cold'' and ``hot'' accretion; e.g.][]{Birnboim2003VirialHaloes, Keres2005, Keres2009a, FG11, vdV11}. 
\cite{Bower2017TheEnd} explained this phenomenon in terms of the suppression of star formation-driven outflows as they lose buoyancy when halo gas becomes hot. 
\cite{Bower2017TheEnd} identified enhanced BH accretion at this halo mass scale as the cause of strong AGN feedback \citep[see also the recent paper by][]{Lapiner2021}. 

Previous studies have analyzed SMBH growth in FIRE simulations \citep[][]{Angles-Alcazar2017GravitationalSimulations, Angles-Alcazar2021CosmologicalHyper-refinement, Catmabacak2022BlackMergers}. 
This study is complementary in its focus on exploring different physical factors that may play a role in explaining when and why SMBH fueling becomes more efficient. 
In this paper, we also analyze a broader range of galaxy masses, including dwarf galaxies. 
In another complementary study, \citet{Wellons2022_AGN} present a broad survey of FIRE simulations including multi-channel AGN feedback.

The plan of this paper is as follows. 
We describe our simulations and analysis methodology in \S \ref{sec:methodology}. 
Our main results are presented in \S \ref{sec:results} and we discuss them in \S \ref{sec:discussion}. 
Our conclusions are summarized in \S \ref{sec:conclusions}.  

\section{Methodology}
\label{sec:methodology}

\subsection{FIRE Simulations}

We analyze a set of FIRE-2 cosmological zoom-in simulations. All simulations were run with the meshless finite mass hydrodynamics code GIZMO \citep{Hopkins2015AMethods}. Details of the FIRE-2 methods and physics are explained in \cite{Hopkins2018FIRE-2Formation}. Throughout, we assume a standard flat $\Lambda$CDM cosmology consistent with recent measurements \citep[][]{Planck2020}.
The simulations include multiple forms of stellar feedback, including feedback from supernovae of Type I and II, stellar winds, photoionization, and radiation pressure on dust grains. 

We examined four massive galaxies, with halo mass $\Mhalo \approx 10^{12.5} \msun$ at $z=2$, which were initially studied by \cite{Angles-Alcazar2017BlackNuclei}. The initial conditions for these galaxies were first presented in \cite{Feldmann2017ColoursNoon}, and they were then re-simulated by \cite{Angles-Alcazar2017BlackNuclei} with a gravitational torque model for black hole accretion, described below. The simulations have mass resolution $\mb = 3.3 \times 10^4 \: \msun$ for baryons and $\mDM = 1.7 \times 10^5 \: \msun$ for dark matter particles, and were run to $z=1$. 
We also examined five `m12' Milky Way-mass galaxies ($\Mhalo = 10^{12} \msun$ at $z=0$) and six `m11' galaxies with $\Mhalo = 10^{11} \msun$ at $z=0$ from the FIRE-2 simulation suite \citep{Wetzel2016RECONCILINGGALAXY,Hopkins2018FIRE-2Formation}, all of which were run to $z=0$. The m11 and m12 simulations have a baryonic mass resolution of $\mb=7100\msun$, with the exceptions of m11b, which has a mass resolution of $\mb=2100\msun$, and m12z, which has a mass resolution of $\mb=4200\msun$. The gravitational softenings for gas particles were again adaptive; the mean softening length in star-forming gas in the m12i simulation was $\epsg = 4.6$ pc \citep{Hopkins2018FIRE-2Formation}. Finally, we analyzed a massive, high-redshift galaxy ($\Mhalo \approx 10^{12.5} \msun$ at $z=5$) first studied by \cite{Ma2020NoSimulationsb}.

We use the Amiga Halo Finder \cite{Knollmann2009Ahf:FINDER} to identify the halo centre, virial mass $\Mhalo$, and virial radius $\Rvir$ of the main halo for each simulation, adopting the virial overdensity definition of \cite{Bryan1998StatisticalComparisons}. We define the stellar mass $\Mstar$ as the total stellar mass within $0.1 \Rvir$.

As the simulations other than those from \cite{Angles-Alcazar2017BlackNuclei} did not include on-the-fly BH accretion calculations, we modeled BH growth in post-processing in all of the simulations, including the massive galaxies for the sake of consistency. \cite{Angles-Alcazar2017BlackNuclei} compared the on-the-fly and post-processing models and found overall good agreement, thus validating the post-processing approach. 
\cite{Catmabacak2022BlackMergers} analyzed how different post-processing assumptions (such as the placement of BHs and the treatment of mergers) affect BH growth in a study focused on massive galaxies.

\subsection{Model for Black Hole Growth} 
\label{sec:BHphysics}
The black hole accretion rate is calculated in post-processing assuming a model in which inflows are driven by gravitational torques, using the same methodology as in in \cite{Angles-Alcazar2017GravitationalSimulations} . Black hole seeds with mass $M_{\rm seed} = 1.4 \times 10^4 \msun$ are introduced in galaxies with stellar masses above $1000 \times M_{\rm seed}$. The black holes are assumed to be located at the centre of the halo, determined using the Amiga Halo Finder as the point in the halo where the combined stellar and dark matter density is highest  \citep{Knollmann2009Ahf:FINDER}. 
The black holes are treated as collisionless particles and allowed to grow through accretion and mergers. The accretion rate is calculated as $\dot{M}_{\rm BH} = (1 - \eta ) \, \dot{M}_{\rm Torque}$, where $\eta = 0.1$ is the constant radiative efficiency. $\dot{M}_{\rm Torque}$ is calculated based on properties of the galaxy within a distance $R_0$ enclosing 256 gas particles (up to a maximum value of $R_0 = 100$ pc$/h$) as 
\begin{equation}
\label{eq:MdotTorque}
\dot{M}_{\rm Torque} = \epsilon _{\rm T} \, f_{\rm d}^{5/2}\, M_{\rm BH,8}^{1/6}\, M_{\rm tot,9} \, R_{0}^{-3/2} \, (1+f_0/f_{\rm gas})^{-1},
\end{equation}
\citep{Hopkins2011AnHoles}, where $\epsilon _{\rm T} = 2.5$ is a normalization factor,  $f_d$ is the disc mass fraction, $\Mtot$ is the total baryonic mass (gas + stars), and
\begin{equation}
\label{eq:f0}
f_0 \approx 0.31 \, f_d^2 \left( \frac{M_d(R_0)}{10^9 \msun} \right) ^{-1/3} 
\end{equation}

Although \citep{Hopkins2011AnHoles} originally tested their gravitational torque estimator on simulations which used a smooth subgrid model for the ISM, based on \cite{SH03_multiphase}, \cite{Hopkins2016_concert} showed that this prescription is also much more accurate at predicting BH accretion than Bondi-like prescriptions in sub-parsec resolution simulations of multiphase galactic nuclei \citep[see also][]{Angles-Alcazar2021CosmologicalHyper-refinement}. 

\cite{Angles-Alcazar2017BlackNuclei} explored several BH growth prescriptions, including models based on the gas mass and free-fall time ($\tff$) near the central BHs, and found that the `kinked' relationship between BH mass and host mass (see Fig. \ref{fig:MBH_Mbulge} for results on this from our analysis) was generic to prescriptions in which the accretion rates were based on the gas content in the inner galactic regions around the black holes. 
Our results are therefore not specific to the gravitational torque model.\footnote{We will see that in the simulations, the onset of accelerated BH fueling correlates with the emergence of steady, thin gas discs (\S \ref{sec:disk_settling}). 
We stress that this result is \emph{not} simply a consequence of the $f_d$ factors in equations (\ref{eq:MdotTorque}) and (\ref{eq:f0}), because our main results are robust to changes in the accretion model in which the disc fraction does not enter (see \S \ref{sec:robust}).}
The normalization of the resulting scaling relation, however, depends on the normalization of the accretion rate prescription (the value of $\epsilon _{\rm T}$ for the gravitational torque model). 
\cite{Angles-Alcazar2017BlackNuclei} also found that the depth of the break in the scaling relation depends on the radius within which the accretion estimator is evaluated in galaxies, such that the break becomes weaker for larger apertures. 
This indicates that the physical conditions in the central regions of galaxies, as opposed to purely galaxy-integrated properties, play an important role in shaping scaling relations \citep[see also][]{Catmabacak2022BlackMergers}.  
We assume that the black holes are at the centres of the simulated galaxies. 
This may result in black holes which can grow more efficiently at early times than in reality. 
For example, \citet{Ma2021_sinking} showed that black hole seeds can take a long time to physically sink to the galaxy centres, especially in early galaxies with clumpy potentials, delaying growth. 
However, this effect goes in the direction of further limiting BH fueling in early galaxies, so it most likely accentuates the relatively inefficient fueling found by neglecting this sinking problem. 

The simulations in this study do not include any form of black hole feedback. 
This allows us to isolate effects on BH fueling that are due to other physics, namely stellar feedback.

\subsection{Comparing the Time of Accelerated BH Growth with Thresholds in Other Physical Properties}
\label{sec:identifying_tgrowth}
\begin{figure}
\begin{center}
\includegraphics[width=.5\textwidth]{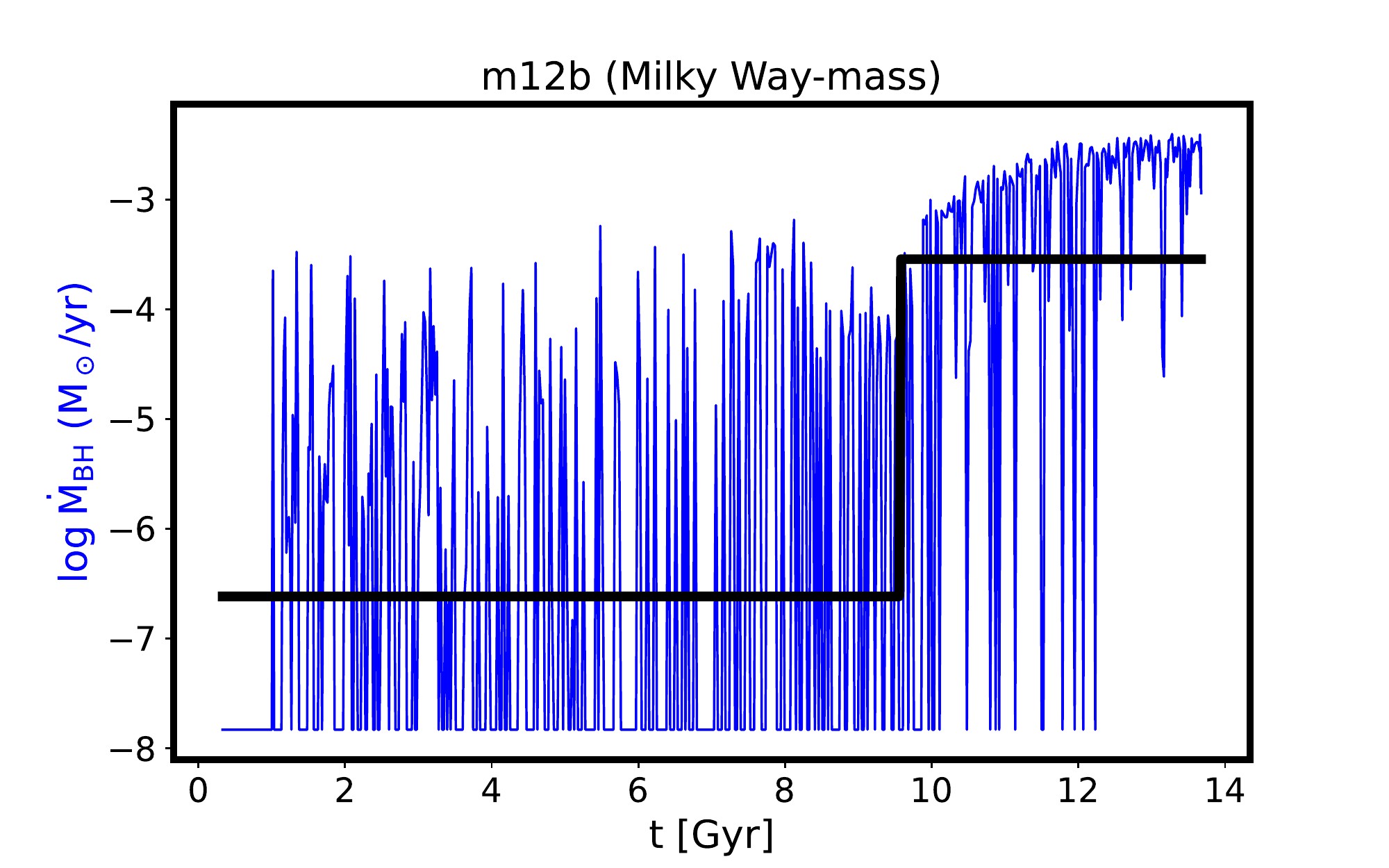}
\end{center}
\caption{Black hole accretion rate (blue) vs.\ age of the universe for one of the galaxies in our sample, m12i, overlaid with the step-function fit (black) used to determine the time of the black hole growth transition, $\tgrowth.$ }
\label{fig:tgrowth_fit}
\end{figure}

To investigate the physical factors that may drive BH growth, we estimate the cosmic time at which the BH accretion rate undergoes a transition to rapid growth in our simulations ($\tgrowth$). Of the sixteen galaxies in our sample, seven did not experience a BH accretion transition (including all of the dwarf galaxies as well as m12z), eight did undergo the transition (including all the massive galaxies---A1, A2, A4, A8, HL09---as well as m12i, m12b, and m12f), and one (m12m) was ambiguous.
For the eight galaxies that experience two phases of black hole growth, we fit a simple step function ($y=A\times ((t-\tgrowth)/(|t-\tgrowth|)+C$) to the BH accretion rate, smoothed with a moving time-average over 300 Myr. An example of this step-function fit for one of our Milky Way-mass galaxies, m12b, is shown in Figure \ref{fig:tgrowth_fit}.
We use the horizontal shift $\tgrowth$ determined by the step function model as the BH growth transition point in subsequent analysis. 

We will analyze how $\tgrowth$ compares with when constant thresholds in other physical properties of the galaxy and its halo are crossed. 
For each property and each galaxy, we calculate $\Delta t = t_{\rm threshold} - \tgrowth$, the time interval between the ``predicted'' onset of accelerated by growth based on different galaxy- or halo-based thresholds and the actual time of accelerated BH growth identified using the simulation data. 
The distributions of these time intervals, which we examine in \S \ref{sec:properties}, can in principle distinguish between ``better'' or ``worse'' predictors. 
For most properties we will consider, we will find the threshold value that minimizes the median $\Delta t$. 

An exception is the time $\tbursty$ at which the galactic star formation rate (SFR) transitions from ``bursty'' (order-of-magnitude time-variable) to time-steady. 
This SFR transition, when it occurs, is rather sharp in the FIRE simulations and can be identified from the SFR time-series alone. 
Therefore we do not attempt to ``optimize'' $\tbursty$ but rather for each simulation we determine it following the method used in \cite{Gurvich2022_disk_settling}. 
Specifically, we evaluate the scatter in 300 Myr windows of SFR, $\sigma_\mathrm{300~Myr}\left(\log(\mathrm{SFR})\right)$. 
The time $\tbursty$ is then defined as the earliest time after which $\sigma_\mathrm{300~Myr}\left(\log(\mathrm{SFR})\right)$ is always below 0.3 dex. 
The time of transition from bursty to steady SFR is of interest because it corresponds closely to changes in the properties of the ISM. 
Before $\tbursty$, the ISM is highly dynamic and frequently ejected by bursts of stellar feedback. 
After $\tbursty$, the simulated galaxies tend to sustain a long-lived gas reservoir in the galaxies \citep[e.g.,][]{Stern2021VirializationFeedback,  Pandya2021CharacterizingSimulations, Gurvich2022_disk_settling}.

\subsection{Inner CGM Virialization}
\label{sec:icv}
In this paper, we investigate how a number of physical properties correlate with black hole growth. 
Most of these properties are straightforward to define and evaluate, but the relationship to the virialization of the inner CGM requires some explanation. 

 A classical understanding of hot halo development suggests that the CGM should virialize when the cooling time $\tcool$ for shocked gas becomes greater than the free-fall time, $\tff$ \citep[e.g.,][]{Rees1977, White1978CoreClustering, Birnboim2003VirialHaloes, Fielding2017TheMedium}.
An important point, studied in idealized models by \cite{Stern2020TheHaloes} and in FIRE cosmological simulations by \cite{Stern2021VirializationFeedback, Stern2021NeutralRedshift}, is that CGM virialization typically proceeds from the outside-in because the $\tcool/\tff$ ratio generally increases outward for realistic gas configurations in haloes.\footnote{This is a subtle point. 
Shocks that heat gas to the virial temperature (including accretion shocks and shocks driven by galactic outflows) often propagate from the inside-out. However, after heating, the inner parts of the CGM cool more rapidly, so these tend to be the last to be become stably supported by thermal pressure.} 
Thus, it is only when $\tcool > \tff$ in the inner CGM, i.e. just outside galaxies rather than on scales $\sim R_{\rm vir}$, that the central galaxy comes into immediate causal contact with the virialized and pressurized CGM. 
\cite{Stern2021VirializationFeedback} showed that inner CGM virialization correlates with a transition from highly bursty to steady star formation rates in FIRE simulations. 
Measures of galaxy ``diskiness,'' such as the ratio of ordered rotation to dispersion for HI gas ($V/\sigma)$, also appear to increase around that time \citep[see also other metrics in][]{Gurvich2022_disk_settling}.  
These galaxy properties may play a role in determining the amount of gas available for BHs to accrete, so it is interesting to quantify better how BH fueling may correlate with inner CGM virialization. 
As mentioned in the introduction, \cite{Bower2017TheEnd} showed that BH growth is suppressed by stellar feedback before hot gaseous haloes form in the EAGLE simulations, although they did not explicitly distinguish between virialization in the inner vs. the outer halo. 

\newcommand{\Tc}{T_{\rm c}}

We identify the expected point in which the inner CGM virializes using the methodology developed in \cite{Stern2021VirializationFeedback}.   This transition is expected when the ratio $\tcool/\tff$ for shocked gas at $r=0.1\Rvir$ exceeds $\sim1$.
For each snapshot, we estimate the free-fall time as: 
\begin{equation}\label{e:tff}
 \tff(r) = \frac{\sqrt{2} r}{\vc} 
\end{equation}
 where $\vc=\sqrt{GM(<r)/r}$ is the circular velocity and $M(<r)$ is the mass enclosed within radius $r$.
The cooling time of shocked gas $\tcool$ is calculated for each snapshot via
\begin{equation}\label{e:tcool}
 \tcool\equiv t_{\rm cool}(\Tc,P_{\rm HSE},Z,z)~,
\end{equation}
where $\Tc\equiv0.6\mu m_{\rm p}\vc^2(r)/k_{\rm B}$ is comparable to the virial temperature, 
$P_{\rm HSE}$ is the thermal pressure assuming hydrostatic equilibrium conditions, estimated via the spherically-averaged weight of gas at $0.1 \Rvir-\Rvir$ (eqn.~12 in \citealt{Stern2021VirializationFeedback}), and $Z$ is the average metallicity at $r$. All properties are evaluated at $r=0.1\, \Rvir$ since this is on average modestly outside the circularization radius for a gas spin parameter $\lambda \sim 0.05$, and hence gas discs contribute negligibly to gas properties\footnote{m11b is an exception since it has a gas disc extending beyond 0.1$\Rvir$. 
In this case, we extrapolate the density from a larger radius, as explained in \cite{Stern2021VirializationFeedback}.}
This calculation of $\tcool$ thus approximates the cooling time in a hot, pressure-supported CGM regardless of whether such a `virialized' CGM actually exists, as done in classic idealized studies \citep[e.g.,][]{White1978CoreClustering,Birnboim2003VirialHaloes}. 
Our previous analyses  \citep{Stern2021VirializationFeedback,Stern2021NeutralRedshift, Gurvich2022_disk_settling} indicate that a hot, stable inner CGM indeed forms at the snapshot when $\tcool/\tff$ exceeds $1-4$, as expected by idealized calculations. We refer the reader to these references for further details on how $\tcool$ and $\tff$ are calculated and a discussion of different approaches.

\section{Results}
\label{sec:results}

\subsection{Overview of Trajectories in the $\Mbh-\Mstar$ Plane}
Figure \ref{fig:MBH_Mbulge} shows the mass of the central black holes as a function of host galaxy stellar mass for all simulations in the sample. For all galaxies regardless of final mass, BHs are undermassive at early times relative to local scaling relations.
The black holes of all massive ($>L_{\star}$) and most Milky Way-mass galaxies then undergo a period of rapid growth and then converge towards the observed scaling relation \citep{Greene2016MEGAMASERGALAXIES} and continue growing steadily.
The transition between the two phases of growth occurs when the host galaxies reach a stellar mass $\sim \Mstarcritrough \msun$, and does not appear to be tied to any particular redshift. 
Although the BH masses at the low end of this relation are sensitive to the seed mass chosen for our model ($M_{\rm seed} = 1.4 \times 10^4 \msun$), the overall kinked shape of the BH trajectories in this plane is robust. 

\subsection{Different Predictors of SMBH Growth}
We would like to understand better the factors that correlate with the increased efficiency in BH fueling around the kink in $\Mbh-\Mstar$, so in this section we explore how BH growth correlates with several other physical quantities. 
In section \ref{sec:properties}, we quantify the evolution of BH masses and growth rates as a function of basic properties of the system on different scales, ranging from the dark matter halo to the inner 1 kpc of galaxies.  
In section \ref{sec:CGMphysics}, we consider some additional physical diagnostics and compare how accurately different threshold crossings predict the timing of accelerated BH growth.

\subsubsection{Correlations with Systemic Properties, from the Halo to the Inner Galaxy}
\label{sec:properties}

\begin{figure}
\begin{center}
\includegraphics[width=.5\textwidth]{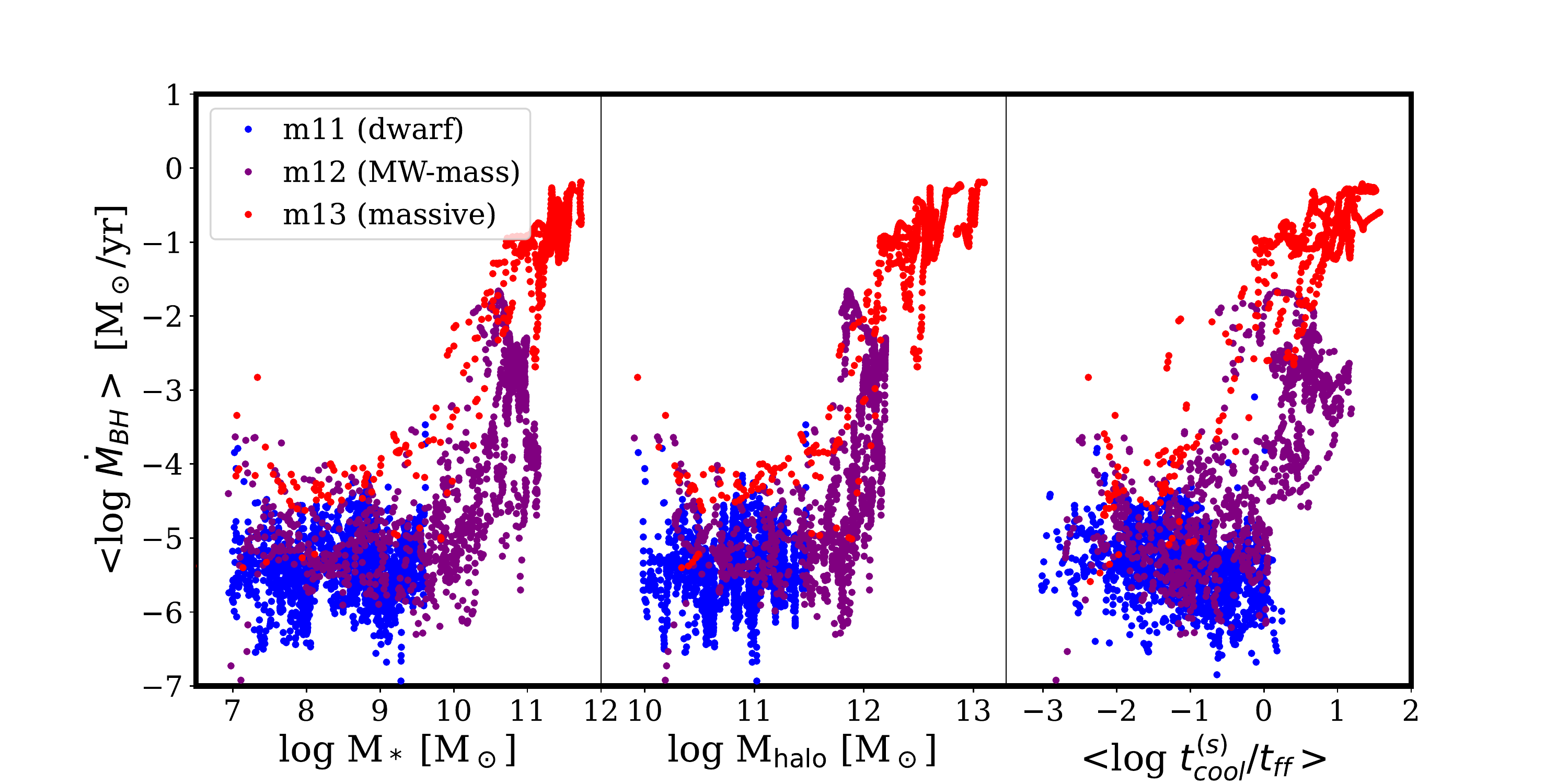}
\end{center}
\caption{Black hole accretion rate vs: $\Mstar$ (left),  $\Mhalo$ (centre), and ratio of $\tcool$ to $\tff$ at $0.1\,\Rvir$ (right). Data is shown for all sixteen simulations in the sample, including galaxies with final halo mass $10^{11} \msun$ (blue), $10^{12} \msun$ (purple), and $> 10^{12} \msun$ (red), and accretion rates and $\tcool/\tff$ values are averaged over 300 Myr. The BH accretion rate increases dramatically above critical values of all three properties.
}
\label{fig:mdot_correlations}
\end{figure}

In Figure \ref{fig:mdot_correlations}, we examine the connection between black hole accretion rate (averaged over 300 Myr) and three other properties of the host halo. The plot includes data from all sixteen galaxies in our sample:  dwarf galaxies, with final halo mass $\sim10^{11} \msun$ (blue), Milky Way-mass galaxies with final halo mass $\sim10^{12} \msun$ (purple), and massive galaxies reaching $\Mhalo \approx 10^{12.5} \msun$ at $z=2$ (red). The left panel plots the BH accretion rate against the stellar mass of the galaxy. We find that the accretion rate remains low (below $10^{-3} \msun$ yr$^{-1}$) and constant below a threshold of around $\Mstar \sim \Mstarcritrough \msun$, then increases drastically above that value. The centre panel plots the BH accretion rate against the total mass of the halo. Once again, the accretion rate is consistently low below a value of $\Mhalo \sim 10^{12} \msun$, then increases rapidly once $\Mhalo$ is above that threshold. The right panel
 plots the BH accretion rate against the ratio of $\tcool$ to $\tff$ at $0.1\,\Rvir$, an indicator of whether the inner CGM is virialized. The accretion rate increases at high values of this ratio, starting at around $\tcool/\tff \approx 1$.

\begin{figure*}
\begin{center}
\includegraphics[width=\textwidth]{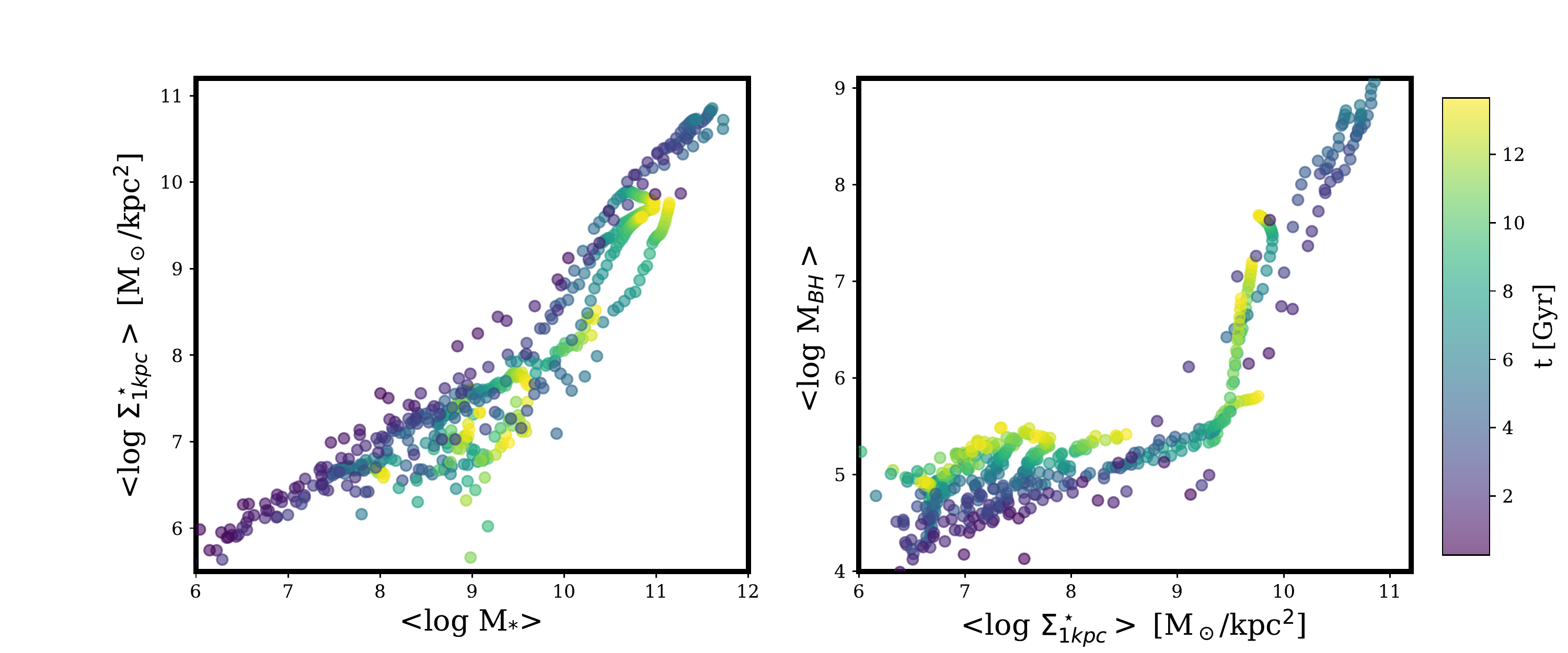}
\end{center}
\caption{The $\SigmaStar$-$\Mstar$ relation (left) and the $\Mbh$-$\SigmaStar$ relation (right) for all galaxies in the simulated sample, averaged over 100 Myr periods and colored by the age of the universe. 
The central stellar surface density increases with stellar mass in the simulations, in qualitative agreement with observations of star-forming galaxies \citep[][]{Chen_Faber2020}. 
The simulations predict that BH mass increases rapidly above a central surface density threshold of about $\SigmaStar \approx 10^{9.5} \msun $kpc$^{-2}$, which may be understood as a signature of gravitational confinement of star formation-driven outflows sustaining a steady gas reservoir from which the nuclear black hole can accrete (see \S \ref{sec:gc})}

\label{fig:Sigma}
\end{figure*}

In Figure \ref{fig:Sigma}, we examine the scaling relations among the stellar surface density in the inner 1 kpc, the stellar mass within $0.1 \Rvir$, and the mass of the central BH, for all sixteen galaxies in the sample. The left panel plots the $\SigmaStar$-$\Mstar$ relation, which shows $\SigmaStar$ increasing monotonically with $\Mstar$.
The right panel plots the $\Mbh$-$\SigmaStar$ relation, which displays a dramatic kink: BH masses increase very rapidly above a surface density threshold of about $\SigmaStar \approx 10^{9.5} \msun$ kpc$^{-2}$. The kink in the $\Mbh$-$\SigmaStar$ relation, like those seen in the $\Mbh$-$\Mstar$, $\MdotBH-\Mstar$ and $\MdotBH-\Mhalo$ relations, suggests that accelerated BH growth is closely tied to changes in physical properties on a range of physical scales. 

To summarize, we find that BH accretion rates are systematically low at low values of $\Mstar$ and $\Mhalo$, then (on average) rise dramatically above threshold values of $\Mstar \sim \Mstarcritrough \msun$ and $\Mhalo \approx 10^{12} \msun$. We see similar behavior in the $\Mbh$-$\SigmaStar$ relation, which is kinked at a critical value of $\SigmaStar \approx 10^{9.5} \msun $ kpc$^{-2}$. 

While many previous studies have found that massive black holes are more directly associated with galactic bulges than with total stellar mass \citep[e.g.,][]{Kormendy2013CoevolutionGalaxies}, our results do not explicitly refer to bulge morphologies. 
We discuss the connection between our results and bulges in more detail in \S \ref{sec:feedback_implications}, but note here that \cite{Hopkins2021WhyWholeb} argued that structures observationally identified as bulges correlate with regions where the stellar surface density exceeds a $\SigmaStar$ threshold similar to the above value.

\subsubsection{Timing Analysis: Is There a ``Best'' Predictor?}
\label{sec:CGMphysics}

\begin{figure*}
\begin{center}
\includegraphics[width=\textwidth]{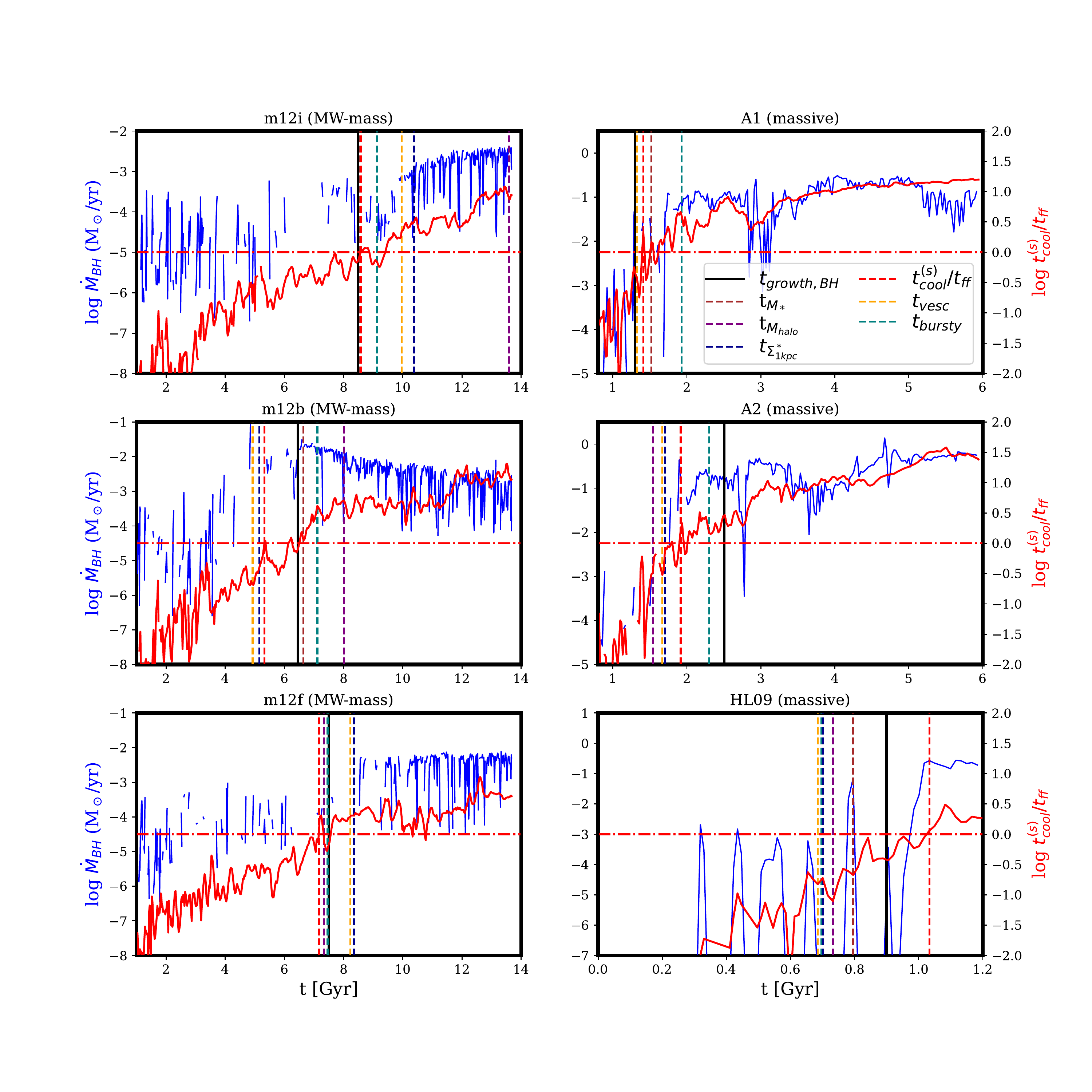}
\end{center}
\caption{Black hole accretion rate (blue, left axis) and ratio of cooling time to free-fall time at 0.1 $\Rvir$ (red, right axis) vs. age of the universe for three Milky Way-mass galaxies (m12i, m12b, m12f), two massive ($\Mhalo \approx 10^{12.5} \msun$ at $z=2$) galaxies (A1, A2), and a massive galaxy at high redshift (HL09). 
In each panel, a $\tcool/\tff$ ratio of unity is marked by a dash-dotted horizontal line. 
The vertical black line marks shows the time of accelerated black hole fueling, $\tgrowth$. The vertical dashed lines represent the points at which the galaxies cross the following best-fitting constant thresholds: $\Mstar = 4 \times 10^{10} \msun$ (teal), $\Mhalo = 1.1 \times 10^{12} \msun$ (yellow), $\tcool/\tff = 1$ (red), $\vesc = 300$ km s$^{-1}$ at a radius of 1 kpc (dark blue) and $\SigmaStar = 10^{9.5} \msun$ kpc$^{-2}$ (purple)
the brown vertical line shows the time at which the galactic star formation rate becomes steady, $\tbursty$. 
Constant thresholds in these different properties and the end of bursty star formation all correlate with accelerated BH black hole fueling.}
\label{fig:timeseries_virialized}
\end{figure*}

\begin{figure*}
\begin{center}
\includegraphics[width=\textwidth]{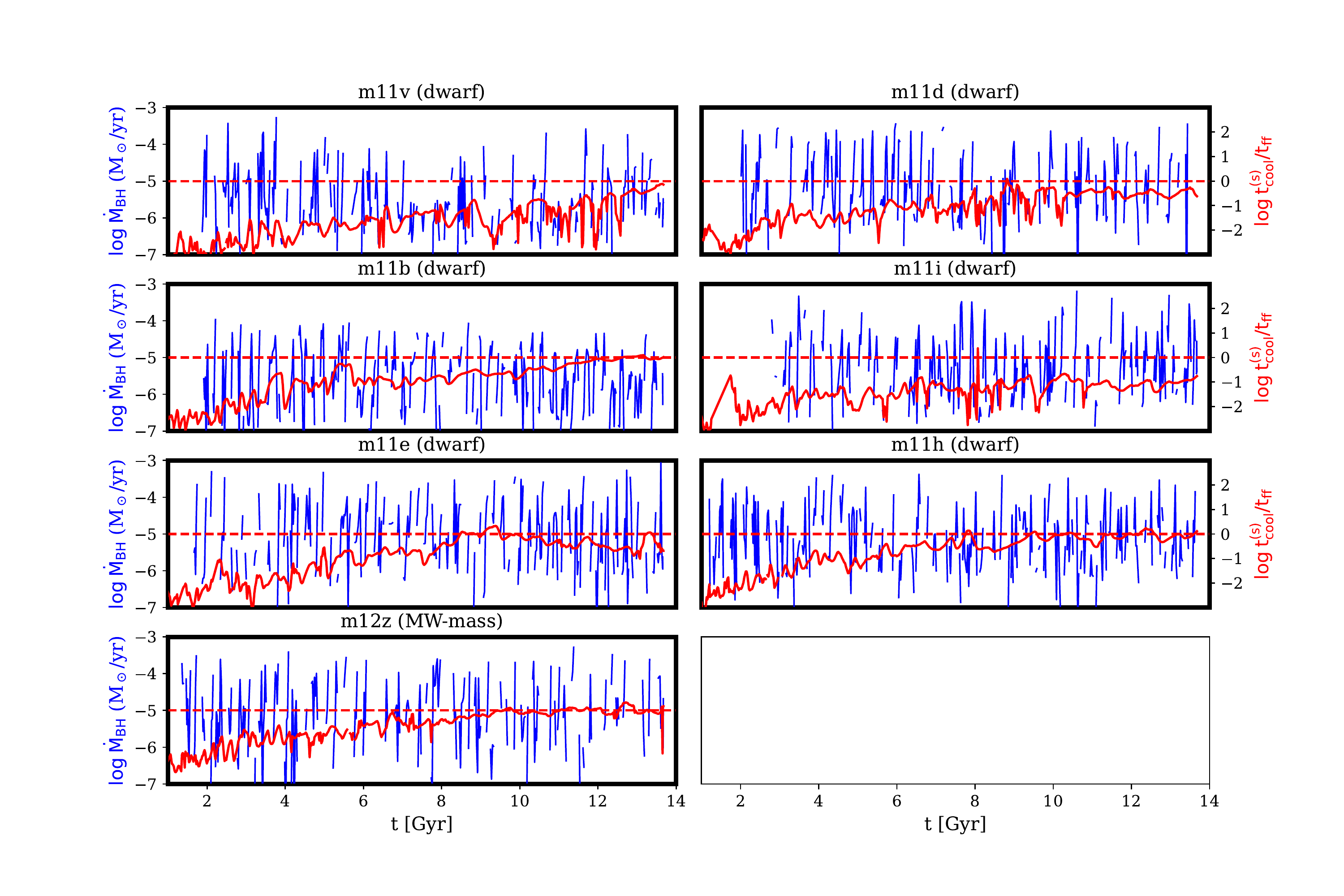}
\end{center}
\caption{Similar to Figure \ref{fig:timeseries_virialized}, but for galaxies that do not experience a transition to accelerated black hole growth.
All analyzed galaxies which did not undergo a transition to efficient black hole growth are shown, including all six m11 galaxies and one m12 galaxy, m12z, the only Milky Way-mass galaxy that does not experience a transition. Though $\tcool/\tff$ approaches or briefly crosses a value of unity in some of these galaxies, it never remains consistently above that value, indicating that the inner CGM does not actually virialize. 
}
\label{fig:timeseries_nonvirialized}
\end{figure*}

It is clear from the above analysis that the efficiency of SMBH feeding correlates with physical properties on a range of scales, from the innermost regions of galaxies (where the BHs are located) to the host dark matter halo. 
However, the causal connections are not obvious. 
In this section, we consider additional physical properties of the host galaxy which may play a role in determining the efficiency of BH feeding. 
Furthermore, we seek to quantify how well thresholds in different galaxy or halo properties can predict the onset of accelerated SMBH growth. 

Figure \ref{fig:timeseries_virialized} shows the BH accretion rate vs. time for six example galaxies that experience a transition in BH growth regime (out of eight in our sample). In each panel, vertical segments compare $\tgrowth$ to the time when the thresholds in $\Mstar$, $\Mhalo$, $\SigmaStar$, and $\tcool/\tff$ are crossed. 
In addition to these quantities analyzed above, the figure also indicates times of crossing thresholds in the escape velocity at 1 kpc from the centre of the galaxy and the time $\tbursty$ at which the galactic SFR transitions from bursty to approximately time-steady. 
The full time evolution of $\tcool/\tff$ at $0.1\,\Rvir$ is also shown in Figure \ref{fig:timeseries_virialized}. 
Figure \ref{fig:timeseries_nonvirialized} shows similar time series, but for lower-mass galaxies that do not experience a transition to more efficient BH fueling. 
None of these galaxies cross the defined thresholds in $\Mstar$, $\Mhalo$, $\vesc$, or $\SigmaStar$. 
Next, we expand on the threshold values favored by the analysis. 

For each quantity, we tested a range of values centered on either the threshold predicted by theory or the approximate thresholds we found earlier. In each case, we optimized the threshold by choosing the value which minimized the median $\Delta t$ for the sample of galaxies.
For the stellar mass and halo mass, there are a priori no specific values at which we expect the efficiency of BH fueling to change drastically. We therefore tested values around $\Mstar \sim \Mstarcritrough \msun$ and $\Mhalo \sim 10^{12} \msun$, which after optimizing yielded thresholds $\Mstar = 4 \times 10^{10} \msun$ and $\Mhalo=1.1 \times 10^{12} \msun$ as the optimal values. 
For $\SigmaStar$, Figure \ref{fig:Sigma} shows a strong kink in the relation between $\Mbh$ and $\SigmaStar$ around $\SigmaStar \sim 10^{9.5} \msun $kpc$^{-2}$, and our optimization process confirmed this value was the optimal threshold. 
As we discuss in \S \ref{sec:mechanisms}, this corresponds to a surface mass density above which stellar feedback is effectively confined.  
For $\tcool/\tff$, the analytic theory suggests virialization of the CGM when the ratio exceeds a value of order unity, so we explored a grid around this value.
The escape velocity at 1 kpc is intended to test a picture in which accelerated BH growth is enabled by the gravitational confinement of supernova-driven outflows \citep[e.g.][]{Dubois2015BlackGrowth,Angles-Alcazar2017BlackNuclei}. 
We tested escape velocities between 100 and 500 km s$^{-1}$ and found that a threshold of 300 km s$^{-1}$ minimized the median $\Delta t$. 

We emphasize that while we optimized the constant threshold for each of the physical quantities discussed in the previous paragraph in order to compare the $\Delta t$ distributions fairly, for $\SigmaStar$ and $\tcool/\tff$ the best-fitting thresholds correspond to values previously derived analytically for the confinement of stellar feedback by gravity \citep[][]{Grudic2020TheFeedback, Hopkins2021WhyWholeb} and for the virialization of the inner CGM \citep[][]{Stern2021VirializationFeedback}. 
The fact that the empirical best-fitting thresholds (optimized over the simulation data set) match these theoretical thresholds suggest causal interpretations.

\begin{figure}
\begin{center}
\includegraphics[width=.5\textwidth]{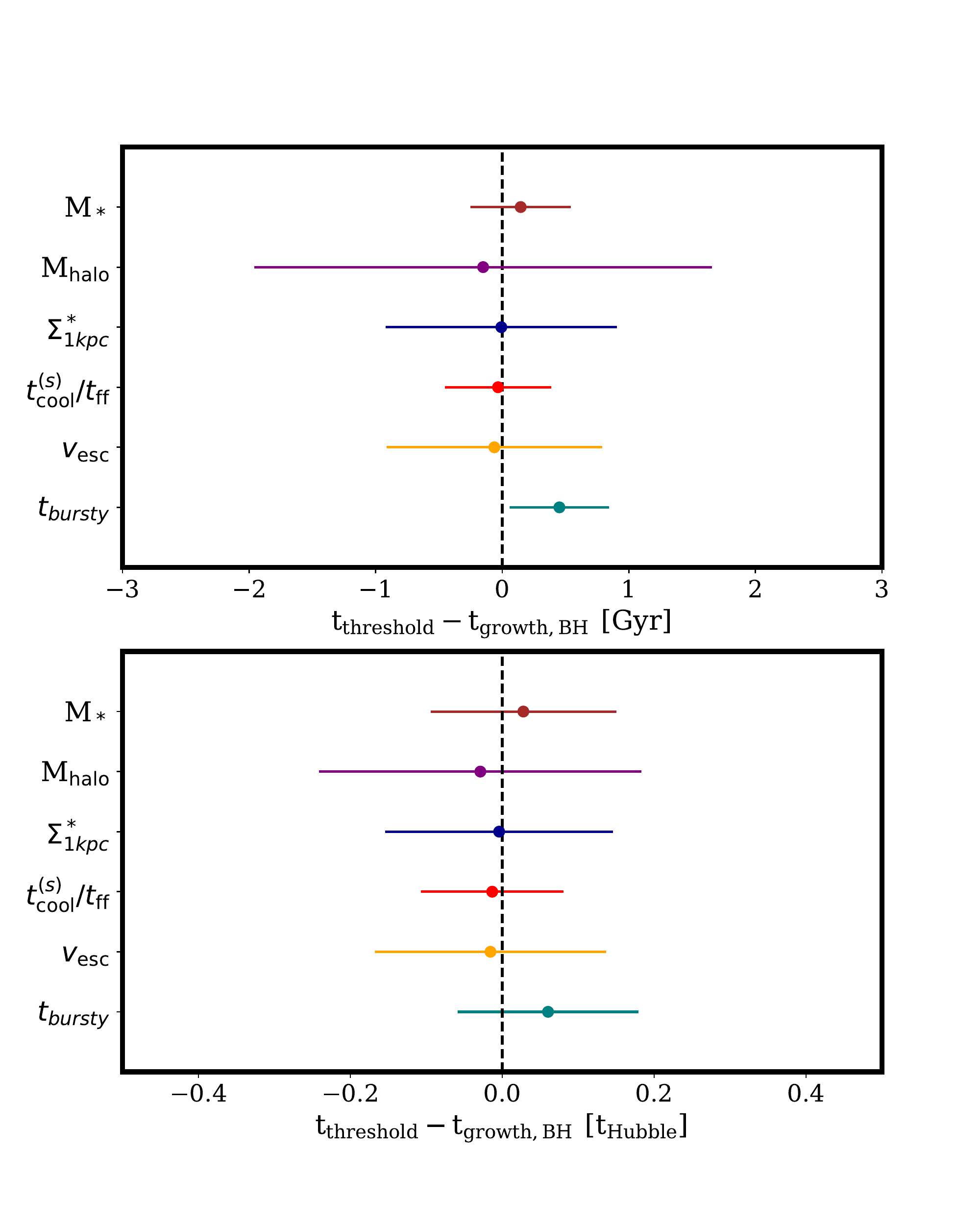}
\end{center} 
\caption{Comparison of the accuracy with which different galaxy properties predict the onset of efficient black hole growth, for the eight galaxies that experience the BH growth transition. For each galaxy, we calculate show the distribution of time differences, in Gyr (top) and in units of the Hubble time (bottom), between the actual cosmic time at which the black hole growth underwent a transition in fueling efficiency and the times at which the following best-fitting constant thresholds are crossed: $\Mstar = 4 \times 10^{10} \msun$ (brown), $\Mhalo = 1.1 \times 10^{12} \msun$ (purple), $\SigmaStar = 10^{9.5} \msun$ kpc$^{-2}$ (dark blue), $\tcool/\tff = 1$ (red), and $\vesc = 300$ km s$^{-1}$ at a radius of 1 kpc (yellow). 
We also show the $\Delta t$ distribution relative to the end transition from bursty to steady star formation, $\tbursty$ (teal). 
Dots indicate the median time difference and the error bars correspond to the 1$\sigma$ width of the distribution in each case. 
The time difference distributions have typical width $\sim 0.1-0.2$ Hubble time, indicating that constant thresholds in these different properties correlate well with accelerated black hole fueling.
}
\label{fig:Delta_t_hists}
\end{figure}

In Figure \ref{fig:Delta_t_hists}, we compare the median and dispersion in $\Delta t$ for the different indicators, relative to the time identified for the BH growth transition.
 The distributions are characterized both in units of absolute time (Gyr; top) and in units of the Hubble time at the BH growth transition, which varies from galaxy to galaxy. 
 The latter is useful because more massive galaxies tend to experience the transitions at higher redshift, i.e.\ when the universe was younger. 
 The median and standard deviation time differences between the BH growth transitions and the times at which the galaxies cross these thresholds are summarized in Table \ref{tab:1}.
 The threshold crossings for the indicators studied are all clustered around the BH growth transition, with standard deviations $\sigma \sim (0.1-0.2) \;\thubble$.  The most precise predictors were $\tcool=\tff$, $\Mstar$, and $\tbursty$, each with
 $1\sigma \leq 0.12 \; \thubble$, while the greatest dispersion in time relative to the BH transition occurred for $\Mhalo$ ($1 \sigma = 0.21 \thubble$). 
 However, since we measure the transitions for only 8 simulations, we do not consider the relative ordering of the threshold crossings or the differences in standard deviations to be statistically significant. 
 The key point is that for all the systemic properties analyzed, constant thresholds are systematically within just $\sim 0.1$ Hubble time of accelerated BH fueling. 
 Absent a physical connection between BH fueling and the galaxy or halo scale properties analyzed, we would expect instead typical offsets in time $\sim  \thubble$, so this is a strong indication that the various galaxy and halo properties we have considered correlate with BH fueling. 
 We stress that this is a highly non-trivial result because of the large separation between the physical scales involved (e.g., BH accretion kernels of radius $\sim 10-100$ pc vs. dark matter haloes of radius $>100$ kpc).
 
\subsection{Robustness of the Analysis with Respect to the Accretion Model}
\label{sec:robust}
As mentioned in section \ref{sec:BHphysics}, \cite{Angles-Alcazar2017BlackNuclei} compared the gravitational torque model of BH accretion to models in which the black hole accretion rate is calculated in post-processing as $\dot{M}_{\rm BH} = SFR(<R_0)/500$ and $\dot{M}_{\rm BH} = \alpha \/ M_{\rm gas} t_{\rm dyn}^{-1}$ where $\alpha$ is a normalization constant $\alpha=10^{-3}-10^{-4}$, and found that the BH growth behavior was similar for all three prescriptions. We verified that the transition between black hole growth regimes occurred approximately simultaneously for all three models, so our results in this section are robust to the details of the BH prescription. 
Additionally, we repeated elements of our analysis on the four massive galaxies from \cite{Angles-Alcazar2017BlackNuclei} with BH accretion modeled on-the-fly during the hydrodynamic simulations (using the gravitational torque model), rather than in post-processing, and obtained similar results. 
 
 \begin{table}[t]
\begin{center}
\begin{tabular}{|c|c|c|c|c|} \hline
Indicator & $\Delta$t$_{1/2}$ [Gyr] & 1$\sigma$ [Gyr] & $\Delta$t$_{1/2}$ [$\thubble$] & 1$\sigma$ [$\thubble$] \\ \hline

$\Mstar$ & 0.15 & 0.40 & 0.03 & 0.12 \\
$\Mhalo$ & -0.15 & 1.81 & -0.03  & 0.21 \\
$\SigmaStar$ & -0.01 & 0.91 & -0.004 & 0.15 \\
$\tcool/\tff$ & -0.03 & 0.42 & -0.01 & 0.09  \\
$\vesc$ & -0.06 & 0.85 & -0.02 & 0.15 \\
$\tbursty$ & 0.45 & 0.39 & 0.06 & 0.12  \\

    \hline
\end{tabular}
\end{center}
\caption{Median ($\Delta$t$_{1/2}$) and standard deviation of the time differences, $\Delta$t $=\tmodel-\tgrowth$, between the transition to accelerated BH growth and the times at which the following best-fitting constant thresholds are crossed: $\Mstar = 4 \times 10^{10} \msun$, $\Mhalo = 1.1 \times 10^{12} \msun$, $\SigmaStar = 10^{9.5} \msun$ kpc$^{-2}$, $\tcool/\tff = 1$, and $\vesc = 300$ km s$^-1$ at a radius of 1 kpc. 
We also show the $\Delta t$ statistics relative to the transition from bursty to steady star formation ($\tbursty$). 
} 
\label{tab:1}
\end{table}

\section{Discussion}
\label{sec:discussion}
In the previous section, we saw that several thresholds in galaxy and halo properties correlate with accelerated BH growth. 
In this section, we suggest possible physical interpretations of the different thresholds, as well as the resulting implications.

\subsection{The Confinement of Stellar Feedback
}
\label{sec:mechanisms}
The two main phases of BH growth are due to a change in how stellar feedback ejects gas (or is unable to eject gas) from the inner galactic regions \citep[][]{Angles-Alcazar2017BlackNuclei}.
We consider two physical mechanisms which can confine and regulate stellar feedback, both of which are consistent with our empirical results, and either or both of which might play a role in driving accelerated black hole growth: gravitational confinement, and pressure confinement by circumgalactic gas. In both cases, we hypothesize that the confinement of star formation-driven outflows contributes to the stabilization of the galactic disc and the buildup of gas available in the central regions to fuel the black hole.

\subsubsection{Gravitational Confinement}
\label{sec:gc}

We found empirically that BH growth increased more rapidly above particular thresholds in $\SigmaStar$ and $\vesc$ at 1 kpc. 
Specifically, the threshold values we found are consistent with the possibility that gravitational confinement of star formation-driven outflows causes changes to the galaxy at that time.
We found that BH mass increases rapidly above a stellar surface density threshold $\SigmaStar \approx 10^{9.5} \msun$~kpc$^{-2} \sim 3000 \msun$ pc$^{-2}$. To within a factor of order unity, this threshold corresponds to the critical surface density in star-forming clouds, $\Sigma_{\rm {tot}} = \SigmaCrit \sim 10^3 \msun $pc$^{-2}$, above which stellar feedback becomes inefficient at ejecting gas. 

This critical surface density can be understood as follows. Star formation efficiency in a gas cloud scales with surface density. This is because star formation regulates itself with stellar feedback: when the momentum-injection from a young stellar population becomes greater than the gravitational force inward on the gas, the cloud will become unbound and its gas expelled. The critical surface density above which stellar feedback becomes inefficient at ejecting gas can be derived by balancing gravity against the momentum of gas driven out by feedback. The rate of momentum outward from a young stellar population is $\langle \dot{p}/m_{\star} \rangle M_{*, young}$, where $\langle\dot{p}/m_{\star}\rangle \sim 6\times10^{-8}$ cm s$^{-2}$ is the characteristic momentum flux output by stellar feedback per unit stellar mass in a young stellar population, and $M_{*, young}$ the mass of young stars \citep[for more details, see ][]{Grudic2018WhenEfficiency, Grudic2020TheFeedback}. The gravitational force inward will simply be $GM_{\rm tot}M_{\rm gas}/R^2$. 
The ratio of the stellar mass formed to the gas mass ejected is 
\begin{equation} \frac{M_{*,young}}{M_{gas}} \approx \frac{GM_{tot}}{<\dot p / m_*>R^2} = \Sigma_{\rm tot}/\SigmaCrit,
\end{equation}
where $\Sigma_{\rm tot} = M_{\rm tot}/\pi R^2$.
This gives us the critical surface density value, $\SigmaCrit \sim \langle \dot{p}/m_{\star} \rangle/G \sim 10^3 \msun $pc$^{-2}$, above which most of the cloud is turned into stars and below which most of the mass is ejected as gas. 
Thus, $\SigmaCrit$ can be interpreted as a critical surface density above which stellar feedback is gravitationally confined.

Interestingly, the numerical value of $\SigmaCrit$ derived from this simple argument is similar to surface density threshold above which we find that BH growth becomes more steady. 
This indicates a possible reason why BH growth becomes more steady (in a time-average sense) above the $\SigmaCrit$ threshold: below this threshold,  stellar feedback efficiently and repeatedly ejects gas in the nuclear regions, whereas above the threshold, a deeper potential well creates a stable, gravitationally bound reservoir of gas which is retained and available to fuel the central BH. \cite{Hopkins2022_Sigmafactor} recently noted this effect specifically in the context of BH fueling \citep[see also][]{Shi2022_hypereddington,Kocevski2017CANDELS:Z2}.

The critical surface density threshold also allows us to understand the value of the threshold in escape velocity which we found to correlate with the onset of more efficient BH fueling. At the time of the transition, we found that $\vesc \approx 300$ km s$^{-1}$ (also evaluated at $R=1$ kpc). 
On a given scale, a total surface density corresponds to a value of the escape velocity. 
Specifically, using $v_{\rm esc}=\sqrt{2 G M_{\rm tot}/R}$ (where $M_{\rm tot}$ is the enclosed mass) and $M_{\rm tot}\approx \pi R^{2} \Sigma_{\rm tot}$, we find
\begin{align}
\vesc & \approx (2 \pi G \Sigma_{\rm tot} R)^{1/2} \\
& \approx 300~{\rm km~s^{-1}} \left( \frac{\Sigma_{\rm tot}}{10^{9.5}~{\rm M_{\odot} kpc^{-2}}} \right)^{1/2} 
\left( \frac{R}{\rm 1 kpc} \right)^{1/2}.
\end{align}

\cite{Dubois2015BlackGrowth} previously obtained a similar result for the critical escape velocity, but based on a different derivation for the velocity expected for winds driven by supernovae in a star-forming clump.

\subsubsection{Pressure Confinement by the Inner CGM}
\label{sec:pc}

Our finding that BH growth is correlated with inner CGM virialization suggests another possible mechanism which could play a role in triggering relevant changes to the central galaxy: pressure confinement. 

We found that the virialization of the inner CGM, as quantified by the cooling time of shocked gas becoming greater than the free-fall time at $0.1 \Rvir$, is also closely correlated in time to the transition to accelerated BH growth. 
This suggests that confinement of star formation-driven outflows by pressure in the inner CGM could play an important role in driving the transition to accelerated BH growth. 

\cite{Stern2021VirializationFeedback} showed that prior to the virialization of the inner CGM, the thermal pressure in the CGM can differ by factors of up to $\sim 100$ in different directions from the halo centre. These pressure fluctuations are dramatically reduced as the inner CGM virializes and becomes homogeneous. Moreover, \cite{Stern2021VirializationFeedback} showed that bursts of supernova-driven outflows in which mass flows out of the galaxy and into the CGM are common prior to ICV but are strongly suppressed as the inner CGM virializes. 
In the scenario suggested by \cite{Stern2021VirializationFeedback}, at early times, the large pressure fluctuations in the CGM provide paths of least resistance for stellar feedback to escape from the galaxy. ``Superbubbles" created by clustered supernovae can therefore disrupt the ISM and prevent the formation of a stable, long-lived gas reservoir \citep[e.g.][]{Martizzi2020GlobalFormation}. After ICV, however, paths of least resistance in the inner CGM effectively close, resulting in more effective confinement of outflows.  

We note that the onset of pressure confinement of star formation-driven outflows upon CGM virialization in FIRE is related to but somewhat different than the mechanism discussed in  \cite{Bower2017TheEnd}, who attributed effective confinement in EAGLE to the loss of outflow buoyancy when the CGM virializes. Prior to virialization the CGM in FIRE does not have a uniform pressure, and thus the concept of buoyancy is not well-defined. 

\subsection{Correlations with Stellar and Halo Masses}
\label{sec:disc_gal_halo_scales}
In the previous section, we discussed confinement mechanisms that can potentially explain why accelerated BH growth is well predicted by thresholds in $\SigmaStar$, $\vesc$, and $\tcool/\tff$. 
Here we address why the BH growth transition also correlates with constant thresholds in $\Mhalo$ and $\Mstar$, and with the end of bursty star formation ($\tbursty$).

We first note that since the relationship between halo mass and stellar mass evolves only weakly with redshift \citep[e.g.,][]{Moster2018EmergeZ10, Behroozi2019UniverseMachine:010}, a constant halo mass threshold maps onto roughly constant stellar mass threshold. 
Therefore, to first order, we only need to explain a constant threshold in one or the other quantity.

In the context of gravitational confinement, the left panel of Figure \ref{fig:Sigma} shows that there is a monotonic and relatively tight (though not one-to-one) relationship between $\SigmaStar$ and $\Mstar$ in our simulated sample. 
Thus, a physical threshold for outflow confinement at a constant value of $\SigmaStar \sim 10^{9.5} \msun$ kpc$^{-2}$ results in a corresponding approximate threshold at $\Mstar \sim 4 \times 10^{10} \msun$. 
Observed star-forming galaxies also have a monotonic average relationship between $\SigmaStar$ and $\Mstar$, although large samples reveal non-negligible dispersion \citep[e.g.,][]{Chen_Faber2020}. 

In the context of pressure confinement, a roughly constant halo mass threshold $\Mhalo \approx 1.1\times10^{12}$ M$_{\odot}$ would follow from the fact that, owing to coincidental cancellations between factors of the cooling function (which enters through $\tcool$) and the properties of dark matter haloes in the standard $\Lambda$CDM cosmology (which enter through both $\tcool$ and $t_{\rm ff}$), the expected halo mass of inner CGM virialization is $\sim 10^{12}$ M$_{\odot}$ and nearly independent of redshift \citep[e.g.,][]{Stern2020TheHaloes}.\footnote{In detail, the halo mass at inner CGM virialization can vary around this value, for example depending on gas metallicity or the depletion of baryons in the CGM due to feedback \citep[e.g.,][]{Stern2020TheHaloes}.} 
Although the emphasis here is on when the inner CGM completes virialization, this is closely related to the large body of theoretical work that has previously shown that the transition between ``cold mode'' and ``hot mode'' accretion occurs at a mass scale that is roughly constant with redshift \citep[e.g.,][]{Birnboim2003VirialHaloes, Keres2005, Keres2009a, FG11, vdV11}.

As discussed further in \S \ref{sec:mechanisms_interaction} and Hopkins et al. (in prep.), CGM virialization correlates strongly with the depth of the gravitational potential, so it is also possible that the different constant thresholds are primarily due to correlations with the depth of the potential.

\subsection{Connection to the Settling of Galactic Discs}
\label{sec:disk_settling}
Here we note that the stabilization of the gas reservoir appears connected to the emergence of steady, thin gas discs in galaxies. 
An indication of this is the fact that the BH accretion rate transition occurs near $\tbursty$, which is when the star formation rate in the ISM transitions from order-of-magnitude time-variable to much more time-steady in FIRE. 
 \cite{Stern2021VirializationFeedback} and \cite{Gurvich2022_disk_settling} showed that the transition from bursty to steady star formation coincides with the rapid ``settling'' of the ISM into a steady, thin disc in the simulations. 
Before $\tbursty$, the ISM is highly dynamic and frequently unbound by bursty stellar feedback. A long-lived, thin gas disc, out of which a thin stellar disc can form, develops after $\tbursty$ \citep[see also][]{Yu2021, Hafen2022_hot_mode}.

\begin{figure*}[h]
\begin{center}
\includegraphics[width=\textwidth]{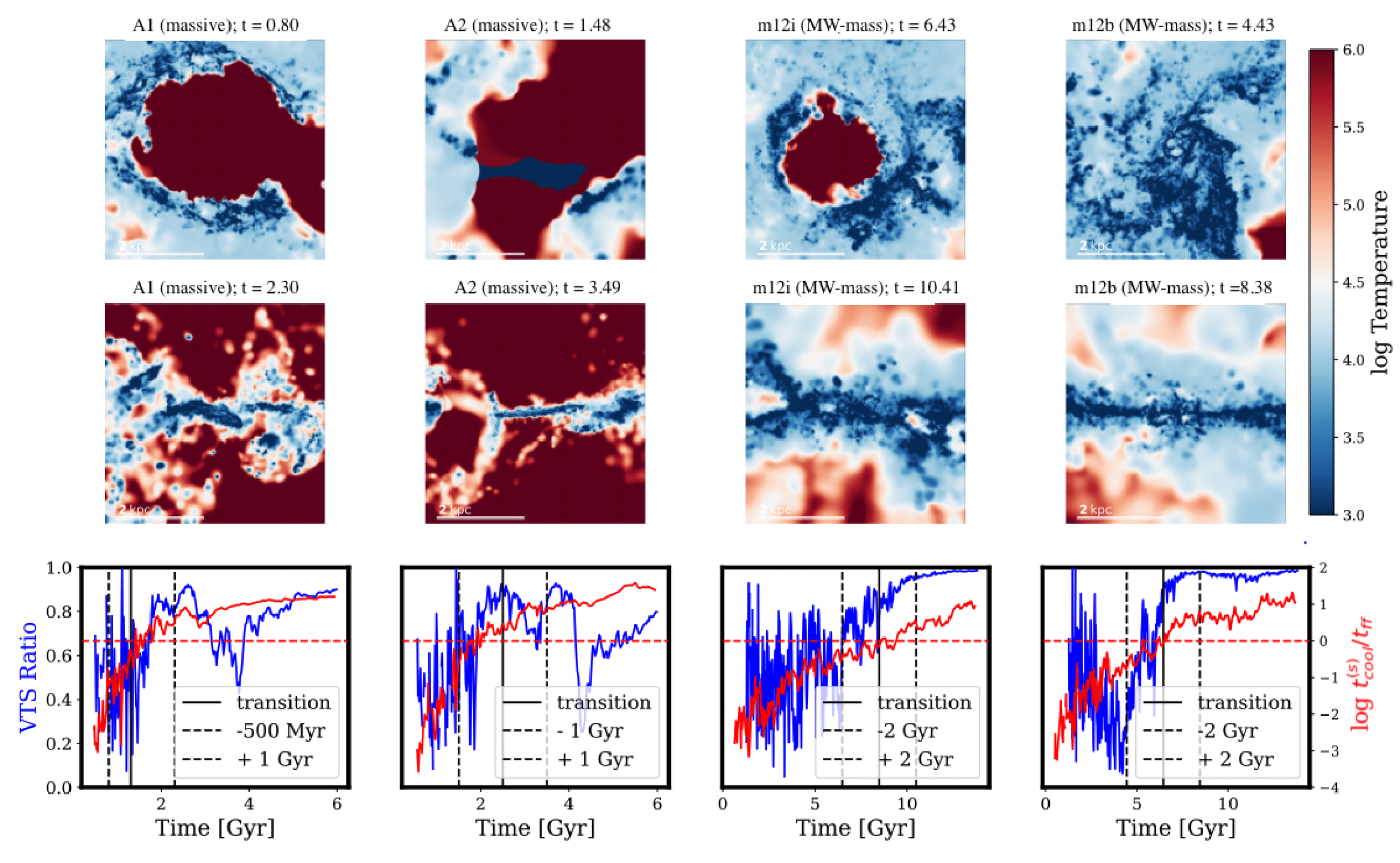}
\end{center}
\caption{ Comparison of the emergence of stable gas discs with inner CGM virialization and BH growth. The top two rows show edge-on gas temperature maps ($\log{T}$ is weighted by mass) before (top) and after (middle) the BH growth  transition. The bottom panels show the quantitative gas disc formation measurement (VTS ratio), as calculated in equation \ref{eqn:VTS}, and ratio of cooling time to free-fall time at 0.1 $\Rvir$ vs. time. We mark the BH growth transition (solid black line) and the snapshots at which the images were taken (dashed black lines). The emergence of stable gas discs correlates with inner CGM virialization and the onset of accelerated BH fueling, though the gas discs in the two massive galaxies (A1 and A2) are later disrupted by mergers. Gas discs are visible in all simulations immediately after the BH growth transition.
}
\label{fig:diskiness}
\end{figure*}

The bursty stellar feedback prior to disc settling frequently clears the galaxy of gas and suppresses the BH's ability to accrete gas from its surroundings. 
The highly intermittent gas distribution around the central BH prior to $\tgrowth$ (the transition to more efficient fueling) is illustrated in Figure \ref{fig:diskiness} for some of our simulations, including two m13 galaxies and two m12 galaxies \citep[the repeated ejection of gas from around the BH at early times was also illustrated in the previous analysis of BH growth in FIRE simulations by][]{Angles-Alcazar2017BlackNuclei}. 
After $\tgrowth$, on the other hand, thin discs are visible in each simulation. 

Figure \ref{fig:diskiness} further illustrates the connections between disc settling, accelerated BH growth, and the virialization of the inner CGM in the bottom row.  
Specifically, the bottom panels show how the ``diskiness" of our galaxies evolves by examining the ``vector-to-scalar'' angular momentum ratio (VTS ratio) of the gas. The VTS ratio is defined as the ratio of the vector to scalar sum of the angular momentum of the resolution elements:
\begin{equation}
        \mathrm{VTS}(r) = \frac{||\sum_{r_i < r} \vec L_i ||}{\sum_{r_i < r} || \vec L_i ||}.\
\label{eqn:VTS}
\end{equation}
Low VTS ratios correspond to random distributions of angular momentum, while near-unity ratios indicate that most of the gas has angular momentum aligned to a single axis, indicating that a stable disc has formed.
Figure \ref{fig:diskiness} compares the evolution of the VTS ratio of the gas within 0.05 $\Rvir$ to the evolution of the $\tcool/\tff$ ratio. 
We find that the formation of thin gas discs generally occurs at roughly the same time as inner CGM virialization ($\tcool/\tff \sim 1$). Note that while this figure focuses on the ICV diagnostic, gravitational confinement of stellar outflows may also play a role in disc settling, as discussed in section \ref{sec:gc} and Hopkins et al. (in prep.). 

In a more detailed analysis of disc settling, \cite{Gurvich2022_disk_settling} find that disc settling in Milky Way-mass galaxies occurs on a time-scale $\sim 1$ Gyr, which is comparable to the width of the $\Delta t$ distribution between the time of accelerated BH growth and the different threshold crossings we have considered (see Fig. \ref{fig:Delta_t_hists}). 
Assuming that disc settling is indeed why the BH's gas supply stabilizes, this suggests that a substantial fraction of the dispersion in $\Delta t$ could be due to the differing finite times it takes for discs to settle.

Before closing this section, we emphasize some important clarifications regarding the connection between disc settling and black hole fueling. 

First, it is the stabilization of the gas reservoir that allows BHs to become more efficiently fed. 
In the FIRE simulations, this corresponds to the formation of a geometrically thin gas disc. 
The stabilization of the gas disc against repeated ejection by stellar feedback is what we term ``disc settling.'' 
However, even in the bursty phase when the ISM is frequently ejected by stellar feedback, there is generally gradual build up of rotational support in the simulations. 
This can be seen in the general increase in the VTS of the gas prior to $\tbursty$  in Figure \ref{fig:diskiness}, and this is discussed at greater length in \cite{Gurvich2022_disk_settling}. 
A stellar distribution that would be identified as a disc, albeit a relatively thick one, can thus appear well before $\tbursty$ \citep[see also][]{Yu2021, Yu2022_born_this_way}. 
Thus, stellar discs generally appear before BHs start growing efficiently. 
This may explain in large part why in observations, BHs often do not correlate well with discs \citep[e.g.,][]{Kormendy2013CoevolutionGalaxies, Greene2016MEGAMASERGALAXIES}. 

Second, we note that the gas supply and star formation rate in the nucleus can continue to experience substantial fluctuations after the galaxy as a whole has stabilized at $\tbursty$ \citep[e.g.,][]{Torrey2017AnNuclei, Orr2021FierySimulations}. 
These nuclear-scale fluctuations are likely important drivers of the short time-scale variability in the BH accretion rates seen even after the time-averaged BH accretion rates become more sustained (see Fig. \ref{fig:timeseries_virialized}). 
However, these nuclear-scale fluctuations do not occur on sufficiently large scales to starve BHs for extended periods of time as is frequent prior to $\tbursty$. 

\subsection{Interaction between Confinement Mechanisms}
\label{sec:mechanisms_interaction}
We found that confinement of stellar feedback by gravity and by gas pressure both correlate tightly with the stabilization of the BH's gas supply. 
We stress that these mechanisms are not mutually exclusive, and may in fact both play a role. 

When the ISM stabilizes and the inner CGM virializes, pressure gradients in the gas on average balance gravity. 
This has been analyzed for the disky ISM of the simulated galaxies in the steady phase \citep[e.g.,][]{Gurvich2020} and is a property of the virialized CGM as well. This suggests that the confining forces from gravity and gas pressure are comparable once the system has reached a quasi-equilibrium.\footnote{There are departures from this equilibrium in the bursty SFR phase, and there can also be later transient departures from equilibrium in the steady phase, e.g. if a galaxy merger triggers a starburst or once AGN feedback becomes important.} 
We note, though, that CGM gas pressure will be most effective at confining hot, volume-filling outflows but may not have a large effect on cold outflows because cold clouds with small cross sections can fly ballistically through dilute hot gas. 
Gravity, on the other hand, acts equally strongly on hot and cold outflows.

The two confining mechanisms may furthermore reinforce each other. 
For example, once gravity confines outflows, the gas densities and metallicities in the inner CGM may be reduced, which could push $\tcool/\tff$ upward and accelerate gas virialization. 
In the opposite direction, if inner CGM virialization begins stabilizing the forming disc, this could accelerate the build up of stellar mass in the inner regions and increase gravitational confinement. 
Such reinforcement between the two confinement mechanisms may play a role in explaining the rapidity of disc settling and the sharpness of the transition to accelerated BH growth.

Although both confinement by gravity and confinement by gas pressure can be important, it is a non-trivial result of our analysis that both mechanisms appear to become important at nearly the same time in the simulated galaxies.  In principle these mechanisms could be decoupled and become important at very different times. 
Here we discuss briefly reasons why the two confinement thresholds may be crossed around the same time. 

\begin{figure}
\begin{center}
\includegraphics[width=.47\textwidth]{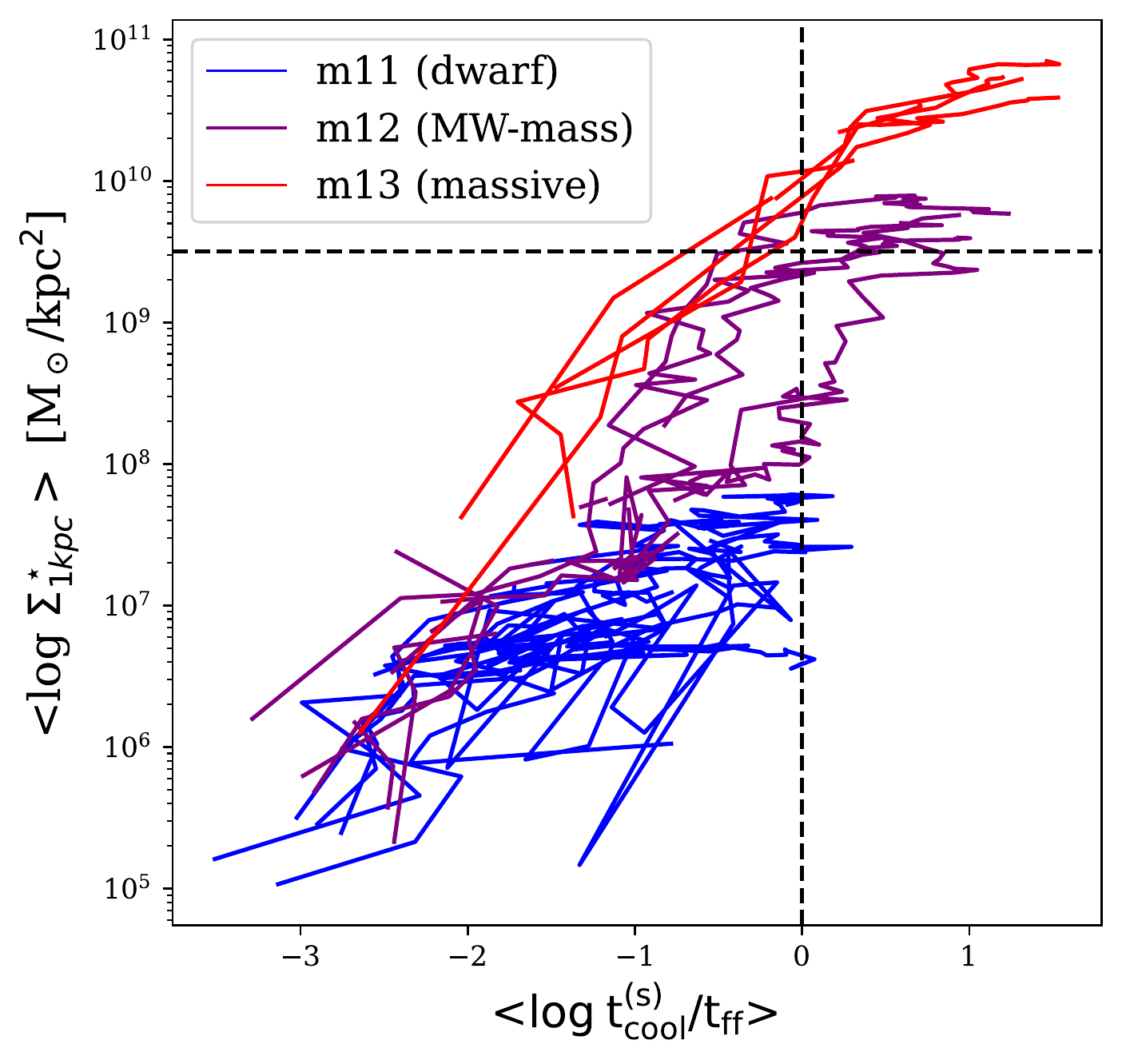}
\end{center}
\caption{ The evolution of the relation between $\SigmaStar$ and $\tcool/\tff$ for all galaxies in the simulated sample, averaged over 300 Myr periods. Colors indicate the final mass of the galaxy. Black dashed lines indicate the empirical threshold values in these quantities which we found best predict the transition to accelerated BH growth ($\SigmaStar = 10^{9.5} \msun \rm{kpc}^{-2}$ and $\tcool/\tff = 1$). We find that galaxies that cross the thresholds generally cross both thresholds at around the same time, with no clear tendency to pass one before the other.}
\label{fig:Sigma1_ICV}
\end{figure}

One possibility is that one threshold crossing causes the other, i.e. the onset of gravitational confinement triggers inner CGM virialization or vice versa (see the previous paragraph for how this could happen). 
However, we find no clear evidence that one threshold crossing systematically precedes the other (compare the $\Delta t$ distributions for $\SigmaStar$ and $\tcool/\tff$ in Fig. \ref{fig:Delta_t_hists}). 
Moreover, it appears that for simulated galaxies that cross the thresholds and experience accelerated BH growth, the galaxies gradually approach each threshold and cross the critical values around the same time, with no clear sign in trajectories in the plane of $\SigmaStar$ vs. $\tcool/\tff$ that crossing of one threshold triggers crossing of the other (see Fig. \ref{fig:Sigma1_ICV}).

Another possibility is that threshold crossings in $\SigmaStar$ and in $\tcool/\tff$ do not cause each other, but generally occur around the same time because of implicit correlations. 
To see why this may be the case, note that both $\SigmaStar$ and $\tcool/\tff$ are functions of the gravitational potential. 
We showed in \S \ref{sec:gc} that $\SigmaStar$ can be expressed in terms of $\vesc$, while \cite{Stern2021VirializationFeedback} showed using analytic approximations that $\tcool/\tff$ can be expressed in terms of $\vc$:
\begin{equation}
\frac{t_{\rm cool}^{\rm (s)}}{t_{\rm ff}} \approx 0.13 v_{100}^{3.4} Z_{1}^{-1} f_{\rm gas}^{-1} (1 + z)^{-1.4} \left( \frac{r}{0.1\Rvir} \right)^{0.5},
\end{equation}
where $v_{\rm 100}=\vc/({\rm 100~km~s^{-1}})$, $Z_{1}$ is the gas metallicity in solar units, and $f_{\rm gas}$ is the gas mass fraction in the CGM relative to the cosmic baryon fraction ($f_{\rm gas}=1$ if the halo contains a cosmic mass fraction $\Omega_{\rm b}/\Omega_{\rm m}$ of baryons, and all these baryons are in the gas phase). 
This equation assumes a fiducial gas profile in the CGM $\rho \propto r^{1.5}$ (see eq. 17 in Stern et al.). 
For a flat rotation curve (i.e., an isothermal potential), the escape velocity $\vesc \approx \sqrt{2} \vc$,\footnote{This relation takes into account only the mass enclosed within the radius of interest, otherwise the escape velocity in an un-truncated isothermal model would be infinite.} so we see that an escape velocity of $300$ km s$^{-1}$ corresponds to a value $\tcool/\tff \sim 1$ at $0.1\Rvir$ for $Z_{1} \sim 1$, $f_{\rm gas}\sim1$, and $z\sim1$. 
Viewed this way, it is not entirely surprising the inner CGM virializes in haloes that have a mass comparable to the haloes hosting galaxies with central potential sufficiently deep to confine stellar feedback gravitationally. 
We leave it to future work to further investigate whether a more precise derivation along these lines (taking into account the exact galaxy and CGM structure) can quantitatively explain the small time differences we find for the two threshold crossings in the simulations.

\subsection{Comparison with Results on Accelerated BH Growth from Other Simulations}
As mentioned in \S \ref{sec:intro}, other studies based on simulations with different sets of subgrid models have found a qualitatively similar transition between BH growth regimes. 
Some previous studies, including \cite{Dubois2015BlackGrowth} and \cite{Angles-Alcazar2017BlackNuclei}, have interpreted this in terms of trapping of star formation-driven outflows once the central potential becomes sufficiently deep. 
Other studies, including \cite{Bower2017TheEnd} and \cite{Lapiner2021}, have emphasized the role of CGM virialization. 
Our analysis shows that both effects turn on around the same time in FIRE, which suggests that both effects could play a role. 
In this section, we compare our results with previous studies. 
Our interpretation in terms of gravitational confinement is broadly similar to previous studies, but there are significant differences in how we interpret the role of CGM virialization, so we focus our discussion on studies that have emphasized this process.

Our hypothesis that inner CGM virialization may play a role in setting the mass scale above which BHs grow more efficiently is related to \cite{Bower2017TheEnd}'s results that the formation of hot haloes triggers accelerated SMBH growth in EAGLE, but there are a few important new elements and distinctions. 
\cite{Bower2017TheEnd} carried out a statistical analysis of a large sample of haloes, and found that accelerated BH growth occurs around a halo mass $M_{\rm h}\sim 10^{12}$ M$_{\odot}$, above which gaseous haloes become hot. 
In their interpretation, stellar feedback-driven outflows limit gas accretion by BHs at early times by keeping central gas densities low, but the stellar outflows are suppressed when haloes reach this mass scale because they are no longer buoyant once the CGM becomes hot. 
This triggers a non-linear response for the BH, which starts to accrete rapidly. 
\cite{McAlpine2018_rapid_growth} expanded on \cite{Bower2017TheEnd}'s analysis and showed that in EAGLE, the onset of rapid BH growth is more accurately associated with a threshold in halo virial temperature than a threshold in halo mass. 
In the pressure confinement scenario (\S \ref{sec:pc}), accelerated BH growth is associated with CGM virialization as in EAGLE, but we more specifically associate it with the emergence of a stable gas disc enabled by the virialization of the \emph{inner} CGM (at $R\sim0.1 \,\Rvir$, just outside the central galaxy), rather than virialization on larger scales, which in general occurs earlier. Additionally, rather than buoyancy being the key factor, we attribute the confinement of outflows post-ICV to the suppression of large pressure fluctuations in virialized gas, and thus the closing of paths of least resistance along which outflows could escape.

The subgrid model differences between the large-volume EAGLE simulations 
 and the FIRE zoom-in simulations analyzed in this paper help us better understand which aspects of the conclusions are sensitive to subgrid modeling. 
 
 For example, the EAGLE simulations use a Bondi-like accretion prescription in which the BH accretion rate scales as $\dot{M}_{\rm BH} \propto M_{\rm BH}^{2}$. 
 Relative to the gravitational torque-driven prescription we adopt, in which $\dot{M}_{\rm BH} \propto M_{\rm BH}^{1/6}$ (see eq. \ref{eq:MdotTorque}), this can suppress BH growth at low $\Mbh$. 
 In \cite{Bower2017TheEnd}'s analytic model of the EAGLE simulations results, the quadratic scaling of the accretion rate with $M_{\rm BH}$ also plays an important role in inducing non-linear growth of SMBHs once they become sufficiently massive.  
 On the other hand, our analysis of FIRE simulations is based on a gravitational torque BH accretion estimator which depends much more weakly on BH mass
 and we have also explored prescriptions in which the accretion rate is completely independent of BH mass 
 \citep[see \S \ref{sec:robust} and][]{Angles-Alcazar2017BlackNuclei}. 
 This shows that non-linearity in the accretion model is not essential to produce a relatively sharp transition between BH accretion fueling regimes. 
 The EAGLE and FIRE simulations also have very different resolutions (baryonic resolution elements of mass $m_{\rm b}\sim 10^{6}$ M$_{\odot}$ in fiducial EAGLE simulations vs. $\sim10^{3}-3\times10^{4}$ M$_{\odot}$ for the zoom-in simulations in this paper), and implement very different subgrid models for the ISM, star formation, stellar feedback, and AGN feedback. 
 In particular, AGN feedback is neglected entirely in the FIRE simulations analyzed in this paper, so it plays no role in our results. 
 This suggests that the transition between BH growth regimes is robust to a broad range of subgrid model variations. 

More recently, \cite{Lapiner2021} analyzed BH growth in the NewHorizon simulation \citep[][]{Dubois2021_NewHorizon}, which zooms onto a spherical volume of comoving radius 10 Mpc and maximum spatial resolution $\sim 40$ pc embedded in the Horizon-AGN simulation box \citep[][]{Dubois2016_HorizonAGN}. 
By $z=0.7$, which is the time that \citet{Lapiner2021} focus on, the NewHorizon volume contains eight galaxies that have grown to a stellar mass greater than $\sim 10^{10}$ $\msun$ (these authors refer to this mass scale and the associated halo mass $M_{\rm h}\sim10^{12}~\msun$ as the ``golden mass''). 
Similar to what we find in FIRE and what has been found in EAGLE, \cite{Lapiner2021} show that BH growth is slow below the ``golden mass'' and accelerates above it. 
This further demonstrates that the existence of two regimes of BH fueling are robust to a range of subgrid model variations. 
As for the comparison with EAGLE, \cite{Lapiner2021}'s interpretation involves similar basic concepts, but is different in detail from ours. 
As in FIRE, the onset of rapid BH growth in NewHorizon is associated with an increase in $\SigmaStar$, which \cite{Lapiner2021} term ``compaction.'' 
Similar to our interpretation, \cite{Lapiner2021} argue that rapid BH growth is enabled when the halo becomes massive enough to retain supernova-driven outflows. 
\cite{Lapiner2021}'s analytic estimates for the ``golden mass'' are based on the confinement of stellar feedback by the gravitational potential and the hot CGM on the scale of the halo.\footnote{\cite{Lapiner2021} present analytic estimates for \emph{two} different critical halo masses, one for the gravitational retention of SN ejecta and one for the stability of virial-radius shocks, which are numerically similar at $z\sim1-2$ and are identified with a single ``golden mass.''}  
We also find that gravity and a hot CGM likely both play important roles, but in our interpretation, it is more precisely when the innermost regions of the CGM virialize that outflows may become pressure confined. 
We also associate the change in BH fueling regime with the settling of the ISM into a long-lived, relatively thin disc, whereas disc settling does not play a prominent role in \cite{Lapiner2021}'s interpretation. 
In \cite{Lapiner2021}'s simulations, the BHs are frequently off-centre at early times but rapidly sink toward to galaxy centres along with the compaction-driven deepening of the gravitational potential and associated dynamical friction. 
In our FIRE analysis, we assume that the BHs are centered at all times (see \S \ref{sec:BHphysics}). 
This shows that the suppression of BH growth at early times is not solely due to mis-centering, although this can be an additional effect \citep[e.g.,][]{Ma2021_sinking, Bahe2021_repositioning}.

\subsection{Implications for Quenching by AGN Feedback, and its Dependence on Stellar Mass and $\SigmaStar$}
\label{sec:feedback_implications}
The insight into BH growth histories given by our results has implications for AGN feedback. 
While there is recent observational evidence for AGN-powered outflows in some dwarf galaxies \citep[e.g.,][]{ManzanoKing2019, Liu2020, Dickey2019AGNGalaxies}, AGN feedback has generally been inferred to affect their host galaxies most strongly at the massive end of the galaxy mass spectrum \citep[e.g.,][]{SD2015_ARAA, NO2017_ARAA}. 
Empirically, it is inferred that the fraction of quenched galaxies increases strongly around $L_{\star}$ \citep[e.g.,][]{Moster2018EmergeZ10, Behroozi2019UniverseMachine:010}. 
Moreover, the incidence of AGN activity increases strongly around this mass scale 
\citep[e.g.,][]{FS14}. 
This corresponds to a mass scale $\Mstar \sim \Mstarcritrough~\msun$ similar to that at which we find BH fueling to become much more efficient in FIRE. 
This is consistent with the hypothesis that AGN feedback is primarily responsible for quenching star formation above this mass scale. 
The physics outlined in the previous sections, namely bursty stellar feedback repeatedly starving BHs at low masses and BH fueling becoming more sustained after the galactic disc stabilizes, could potentially explain the mass scale selected for the quenching of star formation by AGN feedback.

\cite{Chen_Faber2020}, among others, examine observations including galaxy structural properties and argue that the quenching of massive, central galaxies is not fully described by a threshold in stellar mass alone but rather appears to involve (at least) a second parameter. 
Observational evidence also indicates that, among galaxies with similar stellar masses, black hole masses are higher in quiescent galaxies than in star-forming galaxies \citep{Terrazas2016QUIESCENCEGALAXIES}.
There is now a large body of observational studies that show that the central regions of galaxies, as quantified by $\SigmaStar$, are a better determinant of quenching than stellar mass. 
For example, in a survey of galaxies $0.5 \leq z < 0.8$, \cite{Cheung2012TheSurveyb} found that $\SigmaStar$ was the best predictor of whether a galaxy was quenched among a number of other properties analyzed. 
In a sample of SDSS galaxies, \cite{Fang2013AGalaxies} similarly found a close correlation between $\SigmaStar$ and galaxy color, demonstrating that star formation quenches only above a mass-dependent threshold value of $\SigmaStar$.
More recently, \cite{Xu2021CriticalUniverseb} found that central galaxies in the nearby universe quench above $\SigmaStar \sim 10^9 - 10^{9.2} \msun$ kpc$^{-2}$, with the threshold being only weakly dependent on stellar mass. 
Observationally, the connection between central densities and quenching holds at least out to $z\sim3$
\citep[e.g.,][]{vdK2014, Tacchella2015, Barro2017}.

Our results may provide a useful framework to understand the dependence of galaxy quenching on stellar mass and $\SigmaStar$. 
In the picture outlined above, BH growth (and by extension, AGN feedback) is most directly tied to the gas supply in the inner galaxy, which stabilizes when $\SigmaStar$ exceeds a threshold in the range $\sim 10^{9}-10^{9.5}$ $\msun$~kpc$^{-2}$ (\S \ref{sec:gc}). 
Although we found that accelerated BH growth also corresponds to a roughly constant stellar mass threshold $\Mstar \sim 4 \times 10^{10} \msun$ in our simulation sample, this likely arises because of the correlation between $\SigmaStar$ and $\Mstar$ in the simulated galaxies (left panel in Fig. \ref{fig:Sigma}). 
This is because in our physical explanation, the confinement of star formation-driven outflows depends on the depth of the gravitational potential in the inner regions, which depends not only on the total stellar mass of the galaxy but also on how concentrated the mass is around the nucleus. 
With respect to role of the CGM in confining outflows, this also does not in general occur at the same stellar mass in all haloes: the mass at virialization depends, for example, on gas densities and metallicities in the inner CGM and on the exact gravitational potential, which is sensitive to the structural properties of the dark matter halo and the central galaxy \citep[e.g.,][]{Birnboim2003VirialHaloes, Stern2020TheHaloes}.
In future work, it would be interesting to investigate whether these effects can more quantitatively explain the sloped ``ridgeline'' in $\SigmaStar-\Mstar$ that separates star-forming and quenched galaxies observations \citep[][]{Chen_Faber2020}. 
Another effect which could introduce a slope in this ridgeline is the fact that the central surface density depends on the spatial scale on which it is measured, and a radius of 1 kpc is not necessarily the most accurate scale for determining whether outflows are gravitationally confined or not. 
Variations in stellar density profiles with stellar mass could introduce a slope in the $\SigmaStar-\Mstar$ ridgeline if the best indicator of outflow confinement is the surface density evaluated on some other scale. 
This too would be interesting to investigate further in future work.

Virialization of the CGM can also play a key role in determining the effectiveness of AGN feedback for a reason other than the confinement of star formation-driven outflows. 
Several previous studies have highlighted the fact that AGN feedback likely must have a preventive effect in massive haloes, i.e. AGN must prevent too much CGM gas from accreting onto galaxies \citep[e.g.,][]{Croton2006, Bower2006, Keres2009b}. 
When the CGM is hot and virialized, this can be achieved by AGN feedback heating the volume-filling medium. 
On the other hand, prior to CGM virialization the central galaxy is primarily fed by infalling cold streams and cold clouds, which have small geometric cross sections. 
In this regime, it is difficult for AGN feedback (whether be it through collimated jets or wider-angle outflows) to effectively prevent gas infall.

Last, we comment briefly on how our results relate to other concepts commonly discussed in the context of AGN feedback. 
Our results suggest that, on average, BH growth is more efficient and AGN feedback is expected to be stronger above certain thresholds in stellar mass and $\SigmaStar$, and that these thresholds correlate with the settling galaxies into discs. 
In previous literature, both observational and theoretical, BH growth has often been associated with the build up of galaxy bulges, rather than discs, as well as with galaxy mergers \citep[e.g.,][]{Hopkins2008_redEs, Kormendy2013CoevolutionGalaxies}. 
Connections with bulges and a role for galaxy mergers are not necessarily in conflict with our picture. 
First, we stress that BH fueling is highly time variable in our simulations even after the onset of the more efficient fueling phase (see Fig. \ref{fig:timeseries_virialized}). 
This is because even after a relatively stable galactic gas supply has become available, stochastic processes determine exactly when BHs capture some of that gas \citep[e.g.,][]{HH2006_stochastic, Angles-Alcazar2021CosmologicalHyper-refinement}. 
When gas-rich galaxy mergers occur, BH fueling can be rapidly boosted, resulting in a highly energetic burst of AGN feedback \citep[e.g.,][]{Mihos1994_major, DiMatteo2005, Springel2005BlackGalaxies}. 
We suggest that this will occur more commonly after galaxies have grown sufficiently massive to have developed a stable gas reservoir, although it is possible that this also sometimes occurs in lower-mass galaxies between periods of gas ejection by bursty stellar feedback. 
Second, we discussed in \S \ref{sec:disk_settling} how stellar discs are expected to form before the settling of gas into steady, thin discs, and how this will tend to decouple BH masses from the masses of stellar discs identified in observations. 
Third, although we have emphasized the role of disc settling in providing a more steady gas reservoir from which the BH can accrete, this is only the beginning. 
Over time, the central regions can develop a bulge-like morphology, owing either to mergers \citep[][]{Toomre1977_mergers, Hernquist1992_merger_remnants} or to secular processes, such as the in-spiral of massive clumps \citep[][]{Noguchi1999,Dekel2009_analytic}. Finally, as noted by \cite{Hopkins2021WhyWholeb}, observational studies relating properties of bulges to properties of black holes have typically defined those bulges photometrically as excess surface brightness above the surface brightness of the disc. 
\cite{Hopkins2021WhyWholeb} noted that characteristic surface brightness above which bulges are identified corresponds roughly to the critical surface density for gravitational confinement of stellar feedback, $\SigmaCrit$. 
As BH growth becomes efficient above a stellar surface density of approximately $\SigmaCrit$ in our simulations, our findings are  compatible with studies which associate BH growth with stellar bulges.

\subsection{Caveats and Directions for Future Work}
\label{sec:caveats}
In this paper, we have analyzed BH growth by processing zoom-in simulations with a BH accretion rate estimator, and our analysis makes a number of simplifying assumptions. 
We mention in this section some of the ways that our results could be affected, and processes which would be interesting to include in future work.

Most conspicuously, the simulations analyzed in this work neglect AGN feedback entirely. 
This has the benefit of highlighting how the physics of stellar feedback alone can have important implications for the growth of massive BHs. 
Since our key result is that BH growth is relatively slow before the different thresholds in Table \ref{tab:1} are crossed, it is reasonable to hypothesize that including AGN feedback would not significantly change our main results up to the point where those threshold are crossed, and so that the predicted transition between BH fueling regimes would be mostly unaffected. 
In future work, it will be important to extend our analysis to simulations that include a realistic model for AGN feedback, such as the multi-channel AGN feedback models that has recently be incorporated in some FIRE simulations \citep[][]{Wellons2022_AGN}. 
Simulations with AGN feedback would not only allow us to test the effects of AGN feedback on BH growth, but also to study its effects on galaxy quenching and to what extent the stellar mass scale suggested by our analysis for accelerated BH growth is imprinted on galaxy populations. 
We note that, for example, it could take some time after $\tgrowth$ for AGN feedback to strongly affect massive host galaxies 
\citep[e.g.,][]{Davies2022}. 

We also note that while the stellar feedback-driven modulation of BH feeding may play a role in determining the characteristic mass scale of AGN feedback, this is not necessarily the whole story. 
For example, in some of the simulations with AGN feedback analyzed by \cite{Wellons2022_AGN}, models which successfully quench massive galaxies above $L_{\star}$ also tend to ``over-quench'' galaxies right around $L_{\star}$, i.e. produce too many quenched galaxies at this intermediate mass scale relative to observations. 
This suggests that stellar feedback regulation of BH fueling does not necessarily, on its own, limit the effects of AGN feedback to the galaxy masses indicated by observations. 
However, this conclusion may well be sensitive to the exact AGN feedback implementation and other simulation details, such as a resolution, so more work is needed to clarify this issue. 
One possibility is that in addition to a change in feeding efficiency, there is a change in AGN feedback mode involved (such as may be determined by the physics of the accretion flow on the scale of the event horizon). 
Indeed, some cosmological simulations implement a change in AGN feedback mode, e.g. from ``quasar mode'' to ``radio mode'', or from ``thermal'' to ``kinetic,'' in order to reproduce properties of observed galaxy populations \citep[e.g.,][]{Weinberger2018SupermassiveSimulation, Dave2019_SIMBA}.

Related to the fact that AGN feedback is neglected in the simulations, our post-processing analysis assumes a relatively low normalization of the gravitational torque accretion estimator (see \S \ref{sec:BHphysics}). 
The low normalization can be viewed as representing an unresolved, subgrid ``mass loss'' effect that limits how much gas gets accreted by the BH. 
This is a simplistic way of modeling e.g. the effects of some gas turning into stars on the way to the centre, and some gas not making it to the horizon because it is instead ejected in outflows powered by stars or the accretion flow itself. 
The constant normalization was selected such that, by the end of the simulations, the BHs in galaxies sufficient massive to have experienced the efficient growth phase have masses roughly consistent with observed galaxy-black hole scaling relations for massive galaxies \citep[for more on this normalization approach, see][]{AA2013_BHs_without_selfreg, AA2015_torque, Angles-Alcazar2017BlackNuclei, Catmabacak2022BlackMergers}. 
In simulations that explicitly include AGN feedback, it is possible that self-regulation of BH growth would allow or favor a higher normalization of the accretion model. 
If so, BHs could potentially grow more in the early, bursty phase than in our fiducial analysis, which could predict a less pronounced ``jump'' in black hole mass vs. stellar mass above $\Mstar \sim \Mstarcritrough~\msun$. 
Nevertheless, it should still be the case that there is more gas available for accretion after the gas reservoir has stabilized.

We also assume in this work that BHs are seeded and remain at galaxy centres at all times (\S \ref{sec:BHphysics}). 
However, this is not necessarily the case: BH seeds could form away from galaxy centres, and they could take a long time to sink to the nuclear regions. 
Indeed, the time required for a BH to sink to the potential minimum via dynamical friction can be longer than the age of the universe, particularly at high redshift when galaxies tend to be clumpy and are frequently disrupted by mergers and bursts of star formation. 
Even seeds that form in, or reach, the galactic centre may be kicked off-centre by later dynamical interactions \citep{Tremmel2018WanderingHalos,Boldrini2020SubhaloGalaxies}. 
\citet{Ma2021_sinking} integrated BH orbits in early clumpy galaxies from the FIRE simulations, and found that in many circumstances, BH seeds could not be assumed to have reached the galaxy centre. 
\cite{Lapiner2021} found that in the NewHorizon simulation, BHs were driven toward galaxy centres thanks to increased dynamical friction during compaction, though whether this is the case likely depends on the exact seeding model and prescription for unresolved dynamical friction. 
In future work, it would be interesting to further explore how BH growth is affected by the seeding model and accurate orbits in the galactic potential.

\section{Summary and Conclusions}
\label{sec:conclusions}
We analyze BH growth in a sample of sixteen FIRE-2 cosmological zoom-in simulations of galaxies (including dwarf, Milky Way-mass, and more massive galaxies). 
The present analysis neglects AGN feedback, which allows us to isolate effects on BH growth that can be attributed to other processes, including stellar feedback. 

Motivated by previous work, we examine the characteristics of galaxies whose central supermassive black holes experience a transition to rapid growth. We discuss the potential causal relationships at a range of physical scales, as well as the implications for galaxy quenching by AGN feedback. Our analysis yielded the following results:

\begin{enumerate}
    \item We find two phases of BH growth: inefficient growth at early times, and, in some haloes of Milky Way-mass or greater, a transition to increased time-averaged accretion rates at later times. The transition between the two phases of growth occurs when the host galaxies reach a stellar mass $\sim \Mstarcritrough \msun$, a threshold mass which appears to be roughly independent of redshift. 
    These results extend to a broader range of galaxy masses the previous analysis of BH growth in FIRE-2 by \cite{Angles-Alcazar2017BlackNuclei}, who focused on a subset of four of the most massive galaxies. 
    The transition between BH growth regimes found in FIRE is qualitatively similar to that found in other studies using different subgrid models for BH growth \cite[e.g.,][]{Bower2017TheEnd,Lapiner2021}. 
    
    \item To gain insight into the physical factors that drive the transition to accelerated BH growth, we analyze how the timing of the BH accretion transition compares with the crossing of thresholds in different properties of the system, on scales ranging from the galactic nucleus to the dark matter halo, including: the galaxy stellar mass ($\Mstar$), the total halo mass ($\Mhalo$), the stellar surface density within 1 kpc ($\SigmaStar$), the escape velocity at 1 kpc ($\vesc$), and an indicator of whether the inner CGM has virialized ($\tcool/\tff$). 
    Remarkably, constant (i.e., redshift-independent) thresholds in each of these properties correlate relatively tightly with the transition to more efficient BH fueling (within $\sim0.1-0.2$ Hubble time). 
    The best-fitting values for the constant thresholds in these quantities are: $\Mstar = 4\times10^{10}~\msun$, $\Mhalo = 1.1\times10^{12}~\msun$, $\SigmaStar=10^{9.5}~\msun~{\rm kpc^{-2}}$, $\vesc=300~{\rm km~s^{-1}}$, and $\tcool/\tff=1$. 
    Although we optimized the constant threshold for each of these quantities, the best-fitting thresholds found for $\SigmaStar$ and $\tcool/\tff$ correspond to a priori known theoretical thresholds for gravitational confinement of stellar feedback and for the virialization of the inner CGM, respectively, suggesting causal interpretations. 
    
    \item The timing of accelerated BH growth is also consistent with corresponding, on average, with the time $\tbursty$ when the galaxies transition from highly bursty (with order-of-magnitude SFR fluctuations) to much more time-steady star formation rates. 
    In FIRE, the end of bursty star formation coincides with the emergence of stable, thin gas discs \citep[][]{Stern2021VirializationFeedback, Yu2021, Hafen2022_hot_mode, Gurvich2022_disk_settling}. 
    This suggests that the settling of galaxies into thin gas discs with stable gas reservoirs may play an important role in enabling SMBHs to be efficiently fueled.
    
\end{enumerate}    

In these simulations without AGN feedback, the two phases of BH fueling occur because at early times/low masses, bursty stellar feedback ejects gas from the galactic nucleus, which repeatedly starves the BH and makes accretion highly intermittent. 
BHs begin to grow more steadily on average once galaxies develop a stable gas reservoir (though with important short time-scale variability). 
We investigated what drives the gas reservoir to stabilize, and identified two mechanisms that may do so by confining star formation-driven outflows: (1) confinement by central gravity; and (2) confinement by gas pressure in the inner circumgalactic medium. 
Our results with respect to these two confinement mechanisms are the following:
\begin{enumerate}
\setcounter{enumi}{3}
    \item {\bf Gravitational confinement:} We find that BH masses increase dramatically above a threshold of $\SigmaStar \approx 10^{9.5} \msun$ kpc$^{-2}$ (Fig. \ref{fig:Sigma}, right). 
    This critical surface density threshold corresponds to an escape velocity at a radius of 1 kpc of 300 km s$^{-1}$, the value which we found also corresponds to a threshold for accelerated BH growth. 
    A simple force balance argument shows that this surface density threshold corresponds to the critical surface density above which stellar feedback becomes inefficient at ejecting gas. 
    This suggests that confinement of star formation-driven outflows by gravity in the inner galaxy plays an important role in retaining gas to feed BHs.
    
    \item {\bf Pressure confinement:} We also find that accelerated BH growth correlates tightly with the virialization of the inner CGM, as indicated by when the ratio $\tcool/\tff$ exceeds a value of unity at $0.1\Rvir$. 
    (At face value, Figure \ref{fig:Delta_t_hists} shows that the time when $\tcool/\tff$ reaches unity corresponds most tightly with when BHs start growing more efficiently among the thresholds in all properties we have considered, although this may not be statistically significant.) 
    \cite{Stern2021VirializationFeedback} showed that outflows are strongly suppressed following inner CGM virialization, which they interpreted as owing to confinement by hot gas pressure (in contrast to before inner CGM virialization, when order-of-magnitude pressure fluctuations in the CGM allow outflows to more easily leave the central galaxy).
\end{enumerate}    
    
    These confinement mechanisms are not mutually exclusive, and may reinforce each other. 
    It is noteworthy that they appear to ``turn on'' at nearly the same time in the simulated galaxies. 
    We discussed possible connections between the two confinement mechanisms (\S \ref{sec:mechanisms_interaction}), but more work is needed to further disentangle them and/or determine how they are coupled. 
    We noted in particular that, like gravitational confinement, the virialization of the CGM correlates with the depth of the gravitational potential, so the timing coincidences could reflect correlations rather than causation. 
    In the context of gravitational and pressure confinement, the constant thresholds in stellar mass and halo mass follow because of a relationship between $\SigmaStar$ and $\Mstar$ (Fig. \ref{fig:Sigma}, left), and because inner CGM virialization occurs at a roughly constant halo mass $\Mhalo \sim 10^{12}~\msun$ \citep[e.g.,][]{Birnboim2003VirialHaloes, Stern2020TheHaloes}. 
   
    Although we neglected AGN feedback in this study, it is interesting that the stellar mass and halo mass thresholds above which we predict more efficient BH fueling correspond to the $\sim L^{\star}$ mass scale, above which the faction of passive galaxies increases sharply \citep[e.g.,][]{Behroozi2019UniverseMachine:010}. 
    This suggests that the more efficient regime of BH fueling enabled by the trapping of star formation-driven outflows plays a role in determining the strength of AGN feedback.  
    This is related to the suggestion by \cite{Bower2017TheEnd} that the onset of strong AGN feedback is triggered by the failure of stellar feedback. 
    The $\SigmaStar$ threshold we find is also very similar to the surface density threshold that correlates observationally with galaxy quenching \citep[e.g.][and references therein]{Chen_Faber2020}. 
    Our interpretation in terms of confinement of outflows suggests that a $\SigmaStar$ threshold (or possibly a threshold in $\tcool/\tff$, although this is more difficult to observe) is more fundamental than thresholds in stellar mass or halo mass, which follow from correlations.
    Future work should examine more directly how the stellar feedback-regulated phases of BH growth affect AGN feedback, and conversely how the inclusion of self-consistent AGN feedback affects BH fueling.

\section*{Acknowledgements}
We thank Alister Graham for useful comments on scaling relations, and Eliot Quataert for useful comments and discussions. 
LB was supported by the DOE Computer Science Graduate Fellowship through grant DE-SC0020347. 
CAFG was supported by NSF through grants AST-1715216, AST-2108230, and CAREER award AST-1652522; by NASA through grants 17-ATP17-0067 and 21-ATP21-0036; by STScI through grants HST-AR-16124.001-A and HST-GO-16730.016-A; by CXO through grant TM2-23005X; and by the Research Corporation for Science Advancement through a Cottrell Scholar Award.
JS was supported by the Israel Science Foundation (grant No. 2584/21). 
DAA acknowledges support by NSF grants AST-2009687 and AST-2108944, CXO grant TM2-23006X, and Simons Foundation award CCA-1018464. 
SW was supported by an NSF Astronomy and Astrophysics Postdoctoral Fellowship under award AST2001905. 
Support for PFH was provided by NSF Research Grants 1911233, 20009234, 2108318, NSF CAREER grant 1455342, NASA grants 80NSSC18K0562, HST-AR-15800. 
This work was performed in part at Aspen Center for Physics, which is supported by National Science Foundation grant PHY-1607611.
FIRE-2 simulations were generated using Stampede and Stampede 2, via the Extreme Science and Engineering Discovery Environment (XSEDE), supported by NSF grant ACI-1548562, including allocations TG-AST140023, TG-AST140064, TG-AST160048; Blue Waters, supported by the NSF; Frontera, supported by the NSF and TACC, including allocations AST21010 and AST20016;
Pleiades, via the NASA High-End Computing (HEC) Program through the NASA Advanced Supercomputing (NAS) Division at Ames Research Center, including allocations HEC SMD-16-7592, SMD-16-7561, SMD-17-120; and the Quest computing cluster at Northwestern University and the Wheeler cluster at Caltech. 
Support for PFH was provided by NSF Research Grants 1911233, 20009234, 2108318, NSF CAREER grant 1455342, NASA grants 80NSSC18K0562, HST-AR-15800. 
Some of the calculations presented in this work rely on public analysis code developed by Alex Gurvich \citep{Gurvich_abg_python_2021}. 
Some figures were generated with the help of FIRE studio, an open source Python visualization package \citep{Gurvich2022FIRESimulations}. 

\section*{Data Availability}
The data supporting the plots within this article are available on reasonable request to the corresponding author. 
A public version of the GIZMO code is available at \url{http://www.tapir.caltech.edu/~phopkins/Site/GIZMO.html}. Additional data including simulation snapshots, initial conditions, and derived data products are available at \url{http://fire.northwestern.edu/data/}.

\bibliographystyle{yahapj}
\bibliography{references,references_CA, references_additional}

\begin{thebibliography}{}
\providecommand\natexlab[1]{#1}
\providecommand\JournalTitle[1]{#1}

\bibitem[{Angl{\'{e}}s-Alc{\'{a}}zar
  {et~al.}(2017{\natexlab{a}})Angl{\'{e}}s-Alc{\'{a}}zar, Dav{\'{e}},
  Faucher-Gigu{\`{e}}re, {\"{O}}zel, \&
  Hopkins}]{Angles-Alcazar2017GravitationalSimulations}
Angl{\'{e}}s-Alc{\'{a}}zar, D., Dav{\'{e}}, R., Faucher-Gigu{\`{e}}re, C.-A.,
  {\"{O}}zel, F., \& Hopkins, P.~F. 2017{\natexlab{a}},
  \href{http://dx.doi.org/10.1093/mnras/stw2565}{\JournalTitle{Monthly Notices
  of the Royal Astronomical Society}, 464, 2840}

\bibitem[{Angl{\'{e}}s-Alc{\'{a}}zar
  {et~al.}(2017{\natexlab{b}})Angl{\'{e}}s-Alc{\'{a}}zar,
  Faucher-Gigu{\`{e}}re, Quataert, Hopkins, Feldmann, Torrey, Wetzel, \&
  Kere{\v{s}}}]{Angles-Alcazar2017BlackNuclei}
Angl{\'{e}}s-Alc{\'{a}}zar, D., Faucher-Gigu{\`{e}}re, C.-A., Quataert, E.,
  {et~al.} 2017{\natexlab{b}},
  \href{http://dx.doi.org/10.1093/mnrasl/slx161}{\JournalTitle{Monthly Notices
  of the Royal Astronomical Society: Letters}, 472, L109}

\bibitem[{{Angl{\'e}s-Alc{\'a}zar} {et~al.}(2013){Angl{\'e}s-Alc{\'a}zar},
  {{\"O}zel}, \& {Dav{\'e}}}]{AA2013_BHs_without_selfreg}
{Angl{\'e}s-Alc{\'a}zar}, D., {{\"O}zel}, F., \& {Dav{\'e}}, R. 2013,
  \href{http://dx.doi.org/10.1088/0004-637X/770/1/5}{\JournalTitle{\apj}, 770,
  5}

\bibitem[{{Angl{\'e}s-Alc{\'a}zar} {et~al.}(2015){Angl{\'e}s-Alc{\'a}zar},
  {{\"O}zel}, {Dav{\'e}}, {Katz}, {Kollmeier}, \&
  {Oppenheimer}}]{AA2015_torque}
{Angl{\'e}s-Alc{\'a}zar}, D., {{\"O}zel}, F., {Dav{\'e}}, R., {et~al.} 2015,
  \href{http://dx.doi.org/10.1088/0004-637X/800/2/127}{\JournalTitle{\apj},
  800, 127}

\bibitem[{Angl{\'{e}}s-Alc{\'{a}}zar {et~al.}(2021)Angl{\'{e}}s-Alc{\'{a}}zar,
  Quataert, Hopkins, Somerville, Hayward, Faucher-Gigu{\`{e}}re, Bryan,
  Kere{\v{s}}, Hernquist, \&
  Stone}]{Angles-Alcazar2021CosmologicalHyper-refinement}
Angl{\'{e}}s-Alc{\'{a}}zar, D., Quataert, E., Hopkins, P.~F., {et~al.} 2021,
  \href{http://dx.doi.org/10.3847/1538-4357/AC09E8}{\JournalTitle{The
  Astrophysical Journal}, 917, 53}

\bibitem[{{Bah{\'e}} {et~al.}(2021){Bah{\'e}}, {Schaye}, {Schaller}, {Bower},
  {Borrow}, {Chaikin}, {Kugel}, {Nobels}, \&
  {Ploeckinger}}]{Bahe2021_repositioning}
{Bah{\'e}}, Y.~M., {Schaye}, J., {Schaller}, M., {et~al.} 2021,
  \JournalTitle{arXiv e-prints}, arXiv:2109.01489

\bibitem[{{Bailes} {et~al.}(2021){Bailes}, {Berger}, {Brady}, {Branchesi},
  {Danzmann}, {Evans}, {Holley-Bockelmann}, {Iyer}, {Kajita}, {Katsanevas},
  {Kramer}, {Lazzarini}, {Lehner}, {Losurdo}, {L{\"u}ck}, {McClelland},
  {McLaughlin}, {Punturo}, {Ransom}, {Raychaudhury}, {Reitze}, {Ricci},
  {Rowan}, {Saito}, {Sanders}, {Sathyaprakash}, {Schutz}, {Sesana}, {Shinkai},
  {Siemens}, {Shoemaker}, {Thorpe}, {van den Brand}, \& {Vitale}}]{Bailes2021}
{Bailes}, M., {Berger}, B.~K., {Brady}, P.~R., {et~al.} 2021,
  \href{http://dx.doi.org/10.1038/s42254-021-00303-8}{\JournalTitle{Nature
  Reviews Physics}, 3, 344}

\bibitem[{{Baldassare} {et~al.}(2020){Baldassare}, {Dickey}, {Geha}, \&
  {Reines}}]{Baldassare2020}
{Baldassare}, V.~F., {Dickey}, C., {Geha}, M., \& {Reines}, A.~E. 2020,
  \href{http://dx.doi.org/10.3847/2041-8213/aba0c1}{\JournalTitle{\apjl}, 898,
  L3}

\bibitem[{{Barro} {et~al.}(2017){Barro}, {Faber}, {Koo}, {Dekel}, {Fang},
  {Trump}, {P{\'e}rez-Gonz{\'a}lez}, {Pacifici}, {Primack}, {Somerville},
  {Yan}, {Guo}, {Liu}, {Ceverino}, {Kocevski}, \& {McGrath}}]{Barro2017}
{Barro}, G., {Faber}, S.~M., {Koo}, D.~C., {et~al.} 2017,
  \href{http://dx.doi.org/10.3847/1538-4357/aa6b05}{\JournalTitle{\apj}, 840,
  47}

\bibitem[{Behroozi {et~al.}(2019)Behroozi, Wechsler, Hearin, \&
  Conroy}]{Behroozi2019UniverseMachine:010}
Behroozi, P., Wechsler, R.~H., Hearin, A.~P., \& Conroy, C. 2019,
  \href{http://dx.doi.org/10.1093/MNRAS/STZ1182}{\JournalTitle{Monthly Notices
  of the Royal Astronomical Society}, 488, 3143}

\bibitem[{Birnboim \& Dekel(2003)}]{Birnboim2003VirialHaloes}
Birnboim, Y., \& Dekel, A. 2003,
  \href{http://dx.doi.org/10.1046/j.1365-8711.2003.06955.x}{\JournalTitle{Monthly
  Notices of the Royal Astronomical Society}, 345, 349}

\bibitem[{Boldrini {et~al.}(2020)Boldrini, Mohayaee, \&
  Silk}]{Boldrini2020SubhaloGalaxies}
Boldrini, P., Mohayaee, R., \& Silk, J. 2020,
  \href{http://dx.doi.org/10.1093/MNRASL/SLAA043}{\JournalTitle{Monthly Notices
  of the Royal Astronomical Society: Letters}, 495, L12}

\bibitem[{{Bower} {et~al.}(2006){Bower}, {Benson}, {Malbon}, {Helly}, {Frenk},
  {Baugh}, {Cole}, \& {Lacey}}]{Bower2006}
{Bower}, R.~G., {Benson}, A.~J., {Malbon}, R., {et~al.} 2006,
  \href{http://dx.doi.org/10.1111/j.1365-2966.2006.10519.x}{\JournalTitle{\mnras},
  370, 645}

\bibitem[{Bower {et~al.}(2017)Bower, Schaye, Frenk, Theuns, Schaller, Crain, \&
  McAlpine}]{Bower2017TheEnd}
Bower, R.~G., Schaye, J., Frenk, C.~S., {et~al.} 2017,
  \href{http://dx.doi.org/10.1093/mnras/stw2735}{\JournalTitle{Monthly Notices
  of the Royal Astronomical Society}, 465, 32}

\bibitem[{Bryan \& Norman(1998)}]{Bryan1998StatisticalComparisons}
Bryan, G.~L., \& Norman, M.~L. 1998,
  \href{http://dx.doi.org/10.1086/305262/FULLTEXT/}{\JournalTitle{The
  Astrophysical Journal}, 495, 80}

\bibitem[{\c{C}atmabacak {et~al.}(2022)\c{C}atmabacak, Feldmann, Ang{\'{i}}
  Es-Alc{\'{a}}zar, Alc{\'{a}}zar, Andr´, Faucher-Gig{\`{u}}, Hopkins, Duˇ,
  \& Ker{\v{e}}}]{Catmabacak2022BlackMergers}
\c{C}atmabacak, O., Feldmann, R., Ang{\'{i}} Es-Alc{\'{a}}zar, D., {et~al.}
  2022, \href{http://dx.doi.org/10.1093/MNRAS/STAC040}{\JournalTitle{Monthly
  Notices of the Royal Astronomical Society}, 511, 506}

\bibitem[{{Chen} {et~al.}(2020){Chen}, {Faber}, {Koo}, {Somerville}, {Primack},
  {Dekel}, {Rodr{\'\i}guez-Puebla}, {Guo}, {Barro}, {Kocevski}, {van der Wel},
  {Woo}, {Bell}, {Fang}, {Ferguson}, {Giavalisco}, {Huertas-Company}, {Jiang},
  {Kassin}, {Lin}, {Liu}, {Luo}, {Luo}, {Pacifici}, {Pandya}, {Salim}, {Shu},
  {Tacchella}, {Terrazas}, \& {Yesuf}}]{Chen_Faber2020}
{Chen}, Z., {Faber}, S.~M., {Koo}, D.~C., {et~al.} 2020,
  \href{http://dx.doi.org/10.3847/1538-4357/ab9633}{\JournalTitle{\apj}, 897,
  102}

\bibitem[{Cheung {et~al.}(2012)Cheung, Faber, Koo, Dutton, Simard, McGrath,
  Huang, Bell, Dekel, Fang, Salim, Barro, Bundy, Coil, Cooper, Conselice,
  Davis, Dom{\'{i}}nguez, Kassin, Kocevski, Koekemoer, Lin, Lotz, Newman,
  Phillips, Rosario, Weiner, \& Willmer}]{Cheung2012TheSurveyb}
Cheung, E., Faber, S.~M., Koo, D.~C., {et~al.} 2012,
  \href{http://dx.doi.org/10.1088/0004-637X/760/2/131}{\JournalTitle{The
  Astrophysical Journal}, 760, 131}

\bibitem[{Cicone {et~al.}(2014)Cicone, Maiolino, Sturm, Graci{\'{a}}-Carpio,
  Feruglio, Neri, Aalto, Davies, Fiore, Fischer, Garc{\'{i}}a-Burillo,
  Gonz{\'{a}}lez-Alfonso, Hailey-Dunsheath, Piconcelli, \&
  Veilleux}]{Cicone2014MassiveObservations}
Cicone, C., Maiolino, R., Sturm, E., {et~al.} 2014,
  \href{http://dx.doi.org/10.1051/0004-6361/201322464}{\JournalTitle{Astronomy
  {\&} Astrophysics}, 562, A21}

\bibitem[{{Croton} {et~al.}(2006){Croton}, {Springel}, {White}, {De Lucia},
  {Frenk}, {Gao}, {Jenkins}, {Kauffmann}, {Navarro}, \& {Yoshida}}]{Croton2006}
{Croton}, D.~J., {Springel}, V., {White}, S. D.~M., {et~al.} 2006,
  \href{http://dx.doi.org/10.1111/j.1365-2966.2005.09675.x}{\JournalTitle{\mnras},
  365, 11}

\bibitem[{{Dav{\'e}} {et~al.}(2019){Dav{\'e}}, {Angl{\'e}s-Alc{\'a}zar},
  {Narayanan}, {Li}, {Rafieferantsoa}, \& {Appleby}}]{Dave2019_SIMBA}
{Dav{\'e}}, R., {Angl{\'e}s-Alc{\'a}zar}, D., {Narayanan}, D., {et~al.} 2019,
  \href{http://dx.doi.org/10.1093/mnras/stz937}{\JournalTitle{\mnras}, 486,
  2827}

\bibitem[{{Davies} {et~al.}(2022){Davies}, {Pontzen}, \& {Crain}}]{Davies2022}
{Davies}, J.~J., {Pontzen}, A., \& {Crain}, R.~A. 2022,
  \href{http://dx.doi.org/10.1093/mnras/stac1742}{\JournalTitle{\mnras}, 515,
  1430}

\bibitem[{{Dekel} {et~al.}(2009){Dekel}, {Sari}, \&
  {Ceverino}}]{Dekel2009_analytic}
{Dekel}, A., {Sari}, R., \& {Ceverino}, D. 2009,
  \href{http://dx.doi.org/10.1088/0004-637X/703/1/785}{\JournalTitle{\apj},
  703, 785}

\bibitem[{{Di Matteo} {et~al.}(2005){Di Matteo}, {Springel}, \&
  {Hernquist}}]{DiMatteo2005}
{Di Matteo}, T., {Springel}, V., \& {Hernquist}, L. 2005,
  \href{http://dx.doi.org/10.1038/nature03335}{\JournalTitle{\nat}, 433, 604}

\bibitem[{Dickey {et~al.}(2019)Dickey, Geha, Wetzel, \&
  El-Badry}]{Dickey2019AGNGalaxies}
Dickey, C.~M., Geha, M., Wetzel, A., \& El-Badry, K. 2019,
  \href{http://dx.doi.org/10.3847/1538-4357/AB3220}{\JournalTitle{The
  Astrophysical Journal}, 884, 180}

\bibitem[{{Dubois} {et~al.}(2016){Dubois}, {Peirani}, {Pichon}, {Devriendt},
  {Gavazzi}, {Welker}, \& {Volonteri}}]{Dubois2016_HorizonAGN}
{Dubois}, Y., {Peirani}, S., {Pichon}, C., {et~al.} 2016,
  \href{http://dx.doi.org/10.1093/mnras/stw2265}{\JournalTitle{\mnras}, 463,
  3948}

\bibitem[{Dubois {et~al.}(2015)Dubois, Volonteri, Silk, Devriendt, Slyz, \&
  Teyssier}]{Dubois2015BlackGrowth}
Dubois, Y., Volonteri, M., Silk, J., {et~al.} 2015,
  \href{http://dx.doi.org/10.1093/mnras/stv1416}{\JournalTitle{Monthly Notices
  of the Royal Astronomical Society}, 452, 1502}

\bibitem[{{Dubois} {et~al.}(2021){Dubois}, {Beckmann}, {Bournaud}, {Choi},
  {Devriendt}, {Jackson}, {Kaviraj}, {Kimm}, {Kraljic}, {Laigle}, {Martin},
  {Park}, {Peirani}, {Pichon}, {Volonteri}, \& {Yi}}]{Dubois2021_NewHorizon}
{Dubois}, Y., {Beckmann}, R., {Bournaud}, F., {et~al.} 2021,
  \href{http://dx.doi.org/10.1051/0004-6361/202039429}{\JournalTitle{\aap},
  651, A109}

\bibitem[{Fabian(2012)}]{Fabian2012ObservationalFeedback}
Fabian, A.~C. 2012,
  \href{http://dx.doi.org/10.1146/annurev-astro-081811-125521}{\JournalTitle{Annu.
  Rev. Astron. Astrophys}, 50, 455}

\bibitem[{Fang {et~al.}(2013)Fang, Faber, Koo, \& Dekel}]{Fang2013AGalaxies}
Fang, J.~J., Faber, S.~M., Koo, D.~C., \& Dekel, A. 2013,
  \href{http://dx.doi.org/10.1088/0004-637X/776/1/63}{\JournalTitle{Astrophysical
  Journal}, 776, 63}

\bibitem[{{Faucher-Gigu{\`e}re} {et~al.}(2011){Faucher-Gigu{\`e}re},
  {Kere{\v{s}}}, \& {Ma}}]{FG11}
{Faucher-Gigu{\`e}re}, C.-A., {Kere{\v{s}}}, D., \& {Ma}, C.-P. 2011,
  \href{http://dx.doi.org/10.1111/j.1365-2966.2011.19457.x}{\JournalTitle{\mnras},
  417, 2982}

\bibitem[{Feldmann {et~al.}(2017)Feldmann, Quataert, Hopkins,
  Faucher-Gigu{\`{e}}re, \& Kere{\v{s}}}]{Feldmann2017ColoursNoon}
Feldmann, R., Quataert, E., Hopkins, P.~F., Faucher-Gigu{\`{e}}re, C.-A., \&
  Kere{\v{s}}, D. 2017,
  \href{http://dx.doi.org/10.1093/mnras/stx1120}{\JournalTitle{Monthly Notices
  of the Royal Astronomical Society}, 470, 1050}

\bibitem[{Ferrarese \& Merritt(2000)}]{Ferrarese2000AGalaxies}
Ferrarese, L., \& Merritt, D. 2000,
  \href{http://dx.doi.org/10.1086/312838}{\JournalTitle{The Astrophysical
  Journal}, 539, L9}

\bibitem[{Feruglio {et~al.}(2010)Feruglio, Maiolino, Piconcelli, Menci, Aussel,
  Lamastra, \& Fiore}]{Feruglio2010QuasarOutflows}
Feruglio, C., Maiolino, R., Piconcelli, E., {et~al.} 2010,
  \href{http://dx.doi.org/10.1051/0004-6361/201015164}{\JournalTitle{Astronomy
  and Astrophysics}, 518, L155}

\bibitem[{Fielding {et~al.}(2017)Fielding, Quataert, McCourt, \&
  Thompson}]{Fielding2017TheMedium}
Fielding, D., Quataert, E., McCourt, M., \& Thompson, T.~A. 2017,
  \href{http://dx.doi.org/10.1093/MNRAS/STW3326}{\JournalTitle{Monthly Notices
  of the Royal Astronomical Society}, 466, 3810}

\bibitem[{{Fiore} {et~al.}(2017){Fiore}, {Feruglio}, {Shankar}, {Bischetti},
  {Bongiorno}, {Brusa}, {Carniani}, {Cicone}, {Duras}, {Lamastra}, {Mainieri},
  {Marconi}, {Menci}, {Maiolino}, {Piconcelli}, {Vietri}, \&
  {Zappacosta}}]{Fiore2017}
{Fiore}, F., {Feruglio}, C., {Shankar}, F., {et~al.} 2017,
  \href{http://dx.doi.org/10.1051/0004-6361/201629478}{\JournalTitle{\aap},
  601, A143}

\bibitem[{{Fluetsch} {et~al.}(2019){Fluetsch}, {Maiolino}, {Carniani},
  {Marconi}, {Cicone}, {Bourne}, {Costa}, {Fabian}, {Ishibashi}, \&
  {Venturi}}]{Fluetsch2019}
{Fluetsch}, A., {Maiolino}, R., {Carniani}, S., {et~al.} 2019,
  \href{http://dx.doi.org/10.1093/mnras/sty3449}{\JournalTitle{\mnras}, 483,
  4586}

\bibitem[{{F{\"o}rster Schreiber} {et~al.}(2014){F{\"o}rster Schreiber},
  {Genzel}, {Newman}, {Kurk}, {Lutz}, {Tacconi}, {Wuyts}, {Bandara}, {Burkert},
  {Buschkamp}, {Carollo}, {Cresci}, {Daddi}, {Davies}, {Eisenhauer}, {Hicks},
  {Lang}, {Lilly}, {Mainieri}, {Mancini}, {Naab}, {Peng}, {Renzini}, {Rosario},
  {Shapiro Griffin}, {Shapley}, {Sternberg}, {Tacchella}, {Vergani},
  {Wisnioski}, {Wuyts}, \& {Zamorani}}]{FS14}
{F{\"o}rster Schreiber}, N.~M., {Genzel}, R., {Newman}, S.~F., {et~al.} 2014,
  \href{http://dx.doi.org/10.1088/0004-637X/787/1/38}{\JournalTitle{\apj}, 787,
  38}

\bibitem[{Graham \& Scott(2013)}]{Graham2013THECANDIDATES}
Graham, A.~W., \& Scott, N. 2013,
  \href{http://dx.doi.org/10.1088/0004-637X/764/2/151}{\JournalTitle{The
  Astrophysical Journal}, 764, 151}

\bibitem[{Greene {et~al.}(2016)Greene, Seth, Kim, L{\"{a}}sker, Goulding, Gao,
  Braatz, Henkel, Condon, Lo, \& Zhao}]{Greene2016MEGAMASERGALAXIES}
Greene, J.~E., Seth, A., Kim, M., {et~al.} 2016,
  \href{http://dx.doi.org/10.3847/2041-8205/826/2/L32}{\JournalTitle{The
  Astrophysical Journal Letters}, 826, L32}

\bibitem[{Grudi{\'{c}} {et~al.}(2020)Grudi{\'{c}}, Boylan-Kolchin,
  Faucher-Gigu{\`{e}}re, \& Hopkins}]{Grudic2020TheFeedback}
Grudi{\'{c}}, M.~Y., Boylan-Kolchin, M., Faucher-Gigu{\`{e}}re, C.-A., \&
  Hopkins, P.~F. 2020,
  \href{http://dx.doi.org/10.1093/MNRASL/SLAA103}{\JournalTitle{Monthly Notices
  of the Royal Astronomical Society: Letters}, 496, L127}

\bibitem[{Grudi{\'{c}} {et~al.}(2018)Grudi{\'{c}}, Hopkins,
  Faucher-Gigu{\`{e}}re, Quataert, Murray, \&
  Kere{\v{s}}}]{Grudic2018WhenEfficiency}
Grudi{\'{c}}, M.~Y., Hopkins, P.~F., Faucher-Gigu{\`{e}}re, C.-A., {et~al.}
  2018, \href{http://dx.doi.org/10.1093/MNRAS/STY035}{\JournalTitle{Monthly
  Notices of the Royal Astronomical Society}, 475, 3511}

\bibitem[{Gurvich(2021)}]{Gurvich_abg_python_2021}
Gurvich, A. 2021,
  \href{http://dx.doi.org/10.5281/zenodo.5526769}{\JournalTitle{zenodo}}

\bibitem[{Gurvich {et~al.}(2022)Gurvich, {Gurvich}, \&
  B.}]{Gurvich2022FIRESimulations}
Gurvich, A.~B., {Gurvich}, \& B., A. 2022,
  \href{https://ui.adsabs.harvard.edu/abs/2022ascl.soft02006G/abstract}{\JournalTitle{ascl},
  ascl:2202.006}

\bibitem[{{Gurvich} {et~al.}(2020){Gurvich}, {Faucher-Gigu{\`e}re}, {Richings},
  {Hopkins}, {Grudi{\'c}}, {Hafen}, {Wellons}, {Stern}, {Quataert}, {Chan},
  {Orr}, {Kere{\v{s}}}, {Wetzel}, {Hayward}, {Loebman}, \&
  {Murray}}]{Gurvich2020}
{Gurvich}, A.~B., {Faucher-Gigu{\`e}re}, C.-A., {Richings}, A.~J., {et~al.}
  2020, \href{http://dx.doi.org/10.1093/mnras/staa2578}{\JournalTitle{\mnras},
  498, 3664}

\bibitem[{{Gurvich} {et~al.}(2022){Gurvich}, {Stern}, {Faucher-Gigu{\`e}re},
  {Hopkins}, {Wetzel}, {Moreno}, {Hayward}, {Richings}, \&
  {Hafen}}]{Gurvich2022_disk_settling}
{Gurvich}, A.~B., {Stern}, J., {Faucher-Gigu{\`e}re}, C.-A., {et~al.} 2022,
  \JournalTitle{arXiv e-prints}, arXiv:2203.04321

\bibitem[{{Habouzit} {et~al.}(2017){Habouzit}, {Volonteri}, \&
  {Dubois}}]{Habouzit2017}
{Habouzit}, M., {Volonteri}, M., \& {Dubois}, Y. 2017,
  \href{http://dx.doi.org/10.1093/mnras/stx666}{\JournalTitle{\mnras}, 468,
  3935}

\bibitem[{Habouzit {et~al.}(2021)Habouzit, Li, Somerville, Genel, Pillepich,
  Volonteri, Dav{\'{e}}, Rosas-Guevara, McAlpine, Peirani, Hernquist,
  Angl{\'{e}}s-Alc{\'{a}}zar, Reines, Bower, Dubois, Nelson, Pichon, \&
  Vogelsberger}]{Habouzit2021SupermassiveFunction}
Habouzit, M., Li, Y., Somerville, R.~S., {et~al.} 2021,
  \href{http://dx.doi.org/10.1093/MNRAS/STAB496}{\JournalTitle{Monthly Notices
  of the Royal Astronomical Society}, 503, 1940}

\bibitem[{{Hafen} {et~al.}(2022){Hafen}, {Stern}, {Bullock}, {Gurvich}, {Yu},
  {Faucher-Giguere}, {Fielding}, {Angles-Alcazar}, {Quataert}, {Wetzel},
  {Starkenburg}, {Boylan-Kolchin}, {Moreno}, {Feldmann}, {El-Badry}, {Chan},
  {Trapp}, {Keres}, \& {Hopkins}}]{Hafen2022_hot_mode}
{Hafen}, Z., {Stern}, J., {Bullock}, J., {et~al.} 2022, \JournalTitle{arXiv
  e-prints}, arXiv:2201.07235

\bibitem[{{Hernquist}(1992)}]{Hernquist1992_merger_remnants}
{Hernquist}, L. 1992,
  \href{http://dx.doi.org/10.1086/172009}{\JournalTitle{\apj}, 400, 460}

\bibitem[{Hopkins(2015)}]{Hopkins2015AMethods}
Hopkins, P.~F. 2015,
  \href{http://dx.doi.org/10.1093/mnras/stv195}{\JournalTitle{Monthly Notices
  of the Royal Astronomical Society}, 450, 53}

\bibitem[{{Hopkins} {et~al.}(2008){Hopkins}, {Cox}, {Kere{\v{s}}}, \&
  {Hernquist}}]{Hopkins2008_redEs}
{Hopkins}, P.~F., {Cox}, T.~J., {Kere{\v{s}}}, D., \& {Hernquist}, L. 2008,
  \href{http://dx.doi.org/10.1086/524363}{\JournalTitle{\apjs}, 175, 390}

\bibitem[{{Hopkins} \& {Hernquist}(2006)}]{HH2006_stochastic}
{Hopkins}, P.~F., \& {Hernquist}, L. 2006,
  \href{http://dx.doi.org/10.1086/505753}{\JournalTitle{\apjs}, 166, 1}

\bibitem[{Hopkins {et~al.}(2014)Hopkins, Kere{\v{s}}, O{\~{n}}orbe,
  Faucher-Gigu{\`{e}}re, Quataert, Murray, \&
  Bullock}]{Hopkins2014GalaxiesFormation}
Hopkins, P.~F., Kere{\v{s}}, D., O{\~{n}}orbe, J., {et~al.} 2014,
  \href{http://dx.doi.org/10.1093/mnras/stu1738}{\JournalTitle{Monthly Notices
  of the Royal Astronomical Society}, 445, 581}

\bibitem[{Hopkins \& Quataert(2011)}]{Hopkins2011AnHoles}
Hopkins, P.~F., \& Quataert, E. 2011,
  \href{http://dx.doi.org/10.1111/j.1365-2966.2011.18542.x}{\JournalTitle{Monthly
  Notices of the Royal Astronomical Society}, 415, 1027}

\bibitem[{{Hopkins} {et~al.}(2016){Hopkins}, {Torrey}, {Faucher-Gigu{\`e}re},
  {Quataert}, \& {Murray}}]{Hopkins2016_concert}
{Hopkins}, P.~F., {Torrey}, P., {Faucher-Gigu{\`e}re}, C.-A., {Quataert}, E.,
  \& {Murray}, N. 2016,
  \href{http://dx.doi.org/10.1093/mnras/stw289}{\JournalTitle{\mnras}, 458,
  816}

\bibitem[{Hopkins {et~al.}(2021)Hopkins, Wellons, Angles-Alc{\'{a}}zar,
  Faucher-Gigu{\`{e}}re, \& Grudi{\'{c}}}]{Hopkins2021WhyWholeb}
Hopkins, P.~F., Wellons, S., Angles-Alc{\'{a}}zar, D., Faucher-Gigu{\`{e}}re,
  C.~A., \& Grudi{\'{c}}, M.~Y. 2021,
  \href{http://dx.doi.org/10.1093/MNRAS/STAB3458}{\JournalTitle{Monthly Notices
  of the Royal Astronomical Society}, 510, 630}

\bibitem[{{Hopkins} {et~al.}(2022){Hopkins}, {Wellons},
  {Angl{\'e}s-Alc{\'a}zar}, {Faucher-Gigu{\`e}re}, \&
  {Grudi{\'c}}}]{Hopkins2022_Sigmafactor}
{Hopkins}, P.~F., {Wellons}, S., {Angl{\'e}s-Alc{\'a}zar}, D.,
  {Faucher-Gigu{\`e}re}, C.-A., \& {Grudi{\'c}}, M.~Y. 2022,
  \href{http://dx.doi.org/10.1093/mnras/stab3458}{\JournalTitle{\mnras}, 510,
  630}

\bibitem[{Hopkins {et~al.}(2018)Hopkins, Wetzel, Kere{\v{s}},
  Faucher-Gigu{\`{e}}re, Quataert, Boylan-Kolchin, Murray, Hayward,
  Garrison-Kimmel, Hummels, Feldmann, Torrey, Ma, Angl{\'{e}}s-Alc{\'{a}}zar,
  Su, Orr, Schmitz, Escala, Sanderson, Grudi{\'{c}}, Hafen, Kim, Fitts,
  Bullock, Wheeler, Chan, Elbert, \& Narayanan}]{Hopkins2018FIRE-2Formation}
Hopkins, P.~F., Wetzel, A., Kere{\v{s}}, D., {et~al.} 2018,
  \href{http://dx.doi.org/10.1093/mnras/sty1690}{\JournalTitle{Monthly Notices
  of the Royal Astronomical Society}}

\bibitem[{{Kere{\v{s}}} {et~al.}(2009{\natexlab{a}}){Kere{\v{s}}}, {Katz},
  {Dav{\'e}}, {Fardal}, \& {Weinberg}}]{Keres2009b}
{Kere{\v{s}}}, D., {Katz}, N., {Dav{\'e}}, R., {Fardal}, M., \& {Weinberg},
  D.~H. 2009{\natexlab{a}},
  \href{http://dx.doi.org/10.1111/j.1365-2966.2009.14924.x}{\JournalTitle{\mnras},
  396, 2332}

\bibitem[{{Kere{\v{s}}} {et~al.}(2009{\natexlab{b}}){Kere{\v{s}}}, {Katz},
  {Fardal}, {Dav{\'e}}, \& {Weinberg}}]{Keres2009a}
{Kere{\v{s}}}, D., {Katz}, N., {Fardal}, M., {Dav{\'e}}, R., \& {Weinberg},
  D.~H. 2009{\natexlab{b}},
  \href{http://dx.doi.org/10.1111/j.1365-2966.2009.14541.x}{\JournalTitle{\mnras},
  395, 160}

\bibitem[{{Kere{\v{s}}} {et~al.}(2005){Kere{\v{s}}}, {Katz}, {Weinberg}, \&
  {Dav{\'e}}}]{Keres2005}
{Kere{\v{s}}}, D., {Katz}, N., {Weinberg}, D.~H., \& {Dav{\'e}}, R. 2005,
  \href{http://dx.doi.org/10.1111/j.1365-2966.2005.09451.x}{\JournalTitle{\mnras},
  363, 2}

\bibitem[{Knollmann \& Knebe(2009)}]{Knollmann2009Ahf:FINDER}
Knollmann, S.~R., \& Knebe, A. 2009,
  \href{http://dx.doi.org/10.1088/0067-0049/182/2/608}{\JournalTitle{The
  Astrophysical Journal Supplement Series}, 182, 608}

\bibitem[{Kocevski {et~al.}(2017)Kocevski, Barro, Faber, Dekel, Somerville,
  Young, Williams, McIntosh, Georgakakis, Hasinger, Nandra, Civano, Alexander,
  Almaini, Conselice, Donley, Ferguson, Giavalisco, Grogin, Hathi, Hawkins,
  Koekemoer, Koo, McGrath, Mobasher, Gonz{\'{a}}lez, Pforr, Primack, Santini,
  Stefanon, Trump, Wel, Wuyts, \& Yan}]{Kocevski2017CANDELS:Z2}
Kocevski, D.~D., Barro, G., Faber, S.~M., {et~al.} 2017,
  \href{http://dx.doi.org/10.3847/1538-4357/AA8566}{\JournalTitle{The
  Astrophysical Journal}, 846, 112}

\bibitem[{{Kormendy} \& {Ho}(2013)}]{KH2013_ARAA}
{Kormendy}, J., \& {Ho}, L.~C. 2013,
  \href{http://dx.doi.org/10.1146/annurev-astro-082708-101811}{\JournalTitle{\araa},
  51, 511}

\bibitem[{Kormendy \& Ho(2013)}]{Kormendy2013CoevolutionGalaxies}
Kormendy, J., \& Ho, L.~C. 2013,
  \href{http://dx.doi.org/10.1146/annurev-astro-082708-101811}{\JournalTitle{Annual
  Review of Astronomy and Astrophysics}, 51, 511}

\bibitem[{{Lapiner} {et~al.}(2021){Lapiner}, {Dekel}, \&
  {Dubois}}]{Lapiner2021}
{Lapiner}, S., {Dekel}, A., \& {Dubois}, Y. 2021,
  \href{http://dx.doi.org/10.1093/mnras/stab1205}{\JournalTitle{\mnras}, 505,
  172}

\bibitem[{{L{\"a}sker} {et~al.}(2016){L{\"a}sker}, {Greene}, {Seth}, {van de
  Ven}, {Braatz}, {Henkel}, \& {Lo}}]{Lasker2016}
{L{\"a}sker}, R., {Greene}, J.~E., {Seth}, A., {et~al.} 2016,
  \href{http://dx.doi.org/10.3847/0004-637X/825/1/3}{\JournalTitle{\apj}, 825,
  3}

\bibitem[{{Liu} {et~al.}(2020){Liu}, {Veilleux}, {Canalizo}, {Rupke},
  {Manzano-King}, {Bohn}, \& {U}}]{Liu2020}
{Liu}, W., {Veilleux}, S., {Canalizo}, G., {et~al.} 2020,
  \href{http://dx.doi.org/10.3847/1538-4357/abc269}{\JournalTitle{\apj}, 905,
  166}

\bibitem[{{Ma} {et~al.}(2021){Ma}, {Hopkins}, {Ma}, {Angl{\'e}s-Alc{\'a}zar},
  {Faucher-Gigu{\`e}re}, \& {Kelley}}]{Ma2021_sinking}
{Ma}, L., {Hopkins}, P.~F., {Ma}, X., {et~al.} 2021,
  \href{http://dx.doi.org/10.1093/mnras/stab2713}{\JournalTitle{\mnras}},
  \href{http://arxiv.org/abs/2101.02727}{{\sffamily arXiv:2101.02727
  [astro-ph.GA]}}

\bibitem[{Ma {et~al.}(2020)Ma, Quataert, Wetzel, Hopkins,
  Faucher-Gigu{\`{e}}re, \& Kere{\v{s}}}]{Ma2020NoSimulationsb}
Ma, X., Quataert, E., Wetzel, A., {et~al.} 2020,
  \href{http://dx.doi.org/10.1093/MNRAS/STAA2404}{\JournalTitle{Monthly Notices
  of the Royal Astronomical Society}, 498, 2001}

\bibitem[{Magorrian {et~al.}(1998)Magorrian, Tremaine, Richstone, Bender,
  Bower, Dressler, Faber, Gebhardt, Green, Grillmair, Kormendy, \&
  Lauer}]{Magorrian1998TheCenters}
Magorrian, J., Tremaine, S., Richstone, D., {et~al.} 1998,
  \href{http://dx.doi.org/10.1086/300353}{\JournalTitle{The Astronomical
  Journal}, 115, 2285}

\bibitem[{{Manzano-King} {et~al.}(2019){Manzano-King}, {Canalizo}, \&
  {Sales}}]{ManzanoKing2019}
{Manzano-King}, C.~M., {Canalizo}, G., \& {Sales}, L.~V. 2019,
  \href{http://dx.doi.org/10.3847/1538-4357/ab4197}{\JournalTitle{\apj}, 884,
  54}

\bibitem[{Martizzi(2020)}]{Martizzi2020GlobalFormation}
Martizzi, D. 2020,
  \href{http://dx.doi.org/10.1093/MNRAS/STZ3419}{\JournalTitle{Monthly Notices
  of the Royal Astronomical Society}, 492, 79}

\bibitem[{{McAlpine} {et~al.}(2018){McAlpine}, {Bower}, {Rosario}, {Crain},
  {Schaye}, \& {Theuns}}]{McAlpine2018_rapid_growth}
{McAlpine}, S., {Bower}, R.~G., {Rosario}, D.~J., {et~al.} 2018,
  \href{http://dx.doi.org/10.1093/mnras/sty2489}{\JournalTitle{\mnras}, 481,
  3118}

\bibitem[{{Mihos} \& {Hernquist}(1994)}]{Mihos1994_major}
{Mihos}, J.~C., \& {Hernquist}, L. 1994,
  \href{http://dx.doi.org/10.1086/187460}{\JournalTitle{\apjl}, 431, L9}

\bibitem[{Moster {et~al.}(2018)Moster, Naab, \& White}]{Moster2018EmergeZ10}
Moster, B.~P., Naab, T., \& White, S. D.~M. 2018,
  \href{http://dx.doi.org/10.1093/MNRAS/STY655}{\JournalTitle{Monthly Notices
  of the Royal Astronomical Society}, 477, 1822}

\bibitem[{{Naab} \& {Ostriker}(2017)}]{NO2017_ARAA}
{Naab}, T., \& {Ostriker}, J.~P. 2017,
  \href{http://dx.doi.org/10.1146/annurev-astro-081913-040019}{\JournalTitle{\araa},
  55, 59}

\bibitem[{{Nguyen} {et~al.}(2019){Nguyen}, {Seth}, {Neumayer}, {Iguchi},
  {Cappellari}, {Strader}, {Chomiuk}, {Tremou}, {Pacucci}, {Nakanishi},
  {Bahramian}, {Nguyen}, {den Brok}, {Ahn}, {Voggel}, {Kacharov}, {Tsukui},
  {Ly}, {Dumont}, \& {Pechetti}}]{Nguyen2019}
{Nguyen}, D.~D., {Seth}, A.~C., {Neumayer}, N., {et~al.} 2019,
  \href{http://dx.doi.org/10.3847/1538-4357/aafe7a}{\JournalTitle{\apj}, 872,
  104}

\bibitem[{{Noguchi}(1999)}]{Noguchi1999}
{Noguchi}, M. 1999,
  \href{http://dx.doi.org/10.1086/306932}{\JournalTitle{\apj}, 514, 77}

\bibitem[{Orr {et~al.}(2021)Orr, Hatchfield, Battersby, Hayward, Hopkins,
  Wetzel, Benincasa, Loebman, Sormani, \& Klessen}]{Orr2021FierySimulations}
Orr, M.~E., Hatchfield, H.~P., Battersby, C., {et~al.} 2021,
  \href{http://dx.doi.org/10.3847/2041-8213/ABDEBD}{\JournalTitle{The
  Astrophysical Journal Letters}, 908, L31}

\bibitem[{Pandya {et~al.}(2021)Pandya, Fielding, Angl{\'{e}}s-Alc{\'{a}}zar,
  Somerville, Bryan, Hayward, Stern, Kim, Quataert, Forbes,
  Faucher-Gigu{\`{e}}re, Feldmann, Hafen, Hopkins, Kere{\v{s}}, Murray, \&
  Wetzel}]{Pandya2021CharacterizingSimulations}
Pandya, V., Fielding, D.~B., Angl{\'{e}}s-Alc{\'{a}}zar, D., {et~al.} 2021,
  \href{http://dx.doi.org/10.1093/MNRAS/STAB2714}{\JournalTitle{Monthly Notices
  of the Royal Astronomical Society}, 508, 2979}

\bibitem[{{Planck Collaboration} {et~al.}(2020){Planck Collaboration},
  {Aghanim}, {Akrami}, {Ashdown}, {Aumont}, {Baccigalupi}, {Ballardini},
  {Banday}, {Barreiro}, {Bartolo}, {Basak}, {Battye}, {Benabed}, {Bernard},
  {Bersanelli}, {Bielewicz}, {Bock}, {Bond}, {Borrill}, {Bouchet}, {Boulanger},
  {Bucher}, {Burigana}, {Butler}, {Calabrese}, {Cardoso}, {Carron},
  {Challinor}, {Chiang}, {Chluba}, {Colombo}, {Combet}, {Contreras}, {Crill},
  {Cuttaia}, {de Bernardis}, {de Zotti}, {Delabrouille}, {Delouis}, {Di
  Valentino}, {Diego}, {Dor{\'e}}, {Douspis}, {Ducout}, {Dupac}, {Dusini},
  {Efstathiou}, {Elsner}, {En{\ss}lin}, {Eriksen}, {Fantaye}, {Farhang},
  {Fergusson}, {Fernandez-Cobos}, {Finelli}, {Forastieri}, {Frailis},
  {Fraisse}, {Franceschi}, {Frolov}, {Galeotta}, {Galli}, {Ganga},
  {G{\'e}nova-Santos}, {Gerbino}, {Ghosh}, {Gonz{\'a}lez-Nuevo}, {G{\'o}rski},
  {Gratton}, {Gruppuso}, {Gudmundsson}, {Hamann}, {Handley}, {Hansen},
  {Herranz}, {Hildebrandt}, {Hivon}, {Huang}, {Jaffe}, {Jones}, {Karakci},
  {Keih{\"a}nen}, {Keskitalo}, {Kiiveri}, {Kim}, {Kisner}, {Knox},
  {Krachmalnicoff}, {Kunz}, {Kurki-Suonio}, {Lagache}, {Lamarre}, {Lasenby},
  {Lattanzi}, {Lawrence}, {Le Jeune}, {Lemos}, {Lesgourgues}, {Levrier},
  {Lewis}, {Liguori}, {Lilje}, {Lilley}, {Lindholm}, {L{\'o}pez-Caniego},
  {Lubin}, {Ma}, {Mac{\'\i}as-P{\'e}rez}, {Maggio}, {Maino}, {Mandolesi},
  {Mangilli}, {Marcos-Caballero}, {Maris}, {Martin}, {Martinelli},
  {Mart{\'\i}nez-Gonz{\'a}lez}, {Matarrese}, {Mauri}, {McEwen}, {Meinhold},
  {Melchiorri}, {Mennella}, {Migliaccio}, {Millea}, {Mitra},
  {Miville-Desch{\^e}nes}, {Molinari}, {Montier}, {Morgante}, {Moss}, {Natoli},
  {N{\o}rgaard-Nielsen}, {Pagano}, {Paoletti}, {Partridge}, {Patanchon},
  {Peiris}, {Perrotta}, {Pettorino}, {Piacentini}, {Polastri}, {Polenta},
  {Puget}, {Rachen}, {Reinecke}, {Remazeilles}, {Renzi}, {Rocha}, {Rosset},
  {Roudier}, {Rubi{\~n}o-Mart{\'\i}n}, {Ruiz-Granados}, {Salvati}, {Sandri},
  {Savelainen}, {Scott}, {Shellard}, {Sirignano}, {Sirri}, {Spencer},
  {Sunyaev}, {Suur-Uski}, {Tauber}, {Tavagnacco}, {Tenti}, {Toffolatti},
  {Tomasi}, {Trombetti}, {Valenziano}, {Valiviita}, {Van Tent}, {Vibert},
  {Vielva}, {Villa}, {Vittorio}, {Wand elt}, {Wehus}, {White}, {White},
  {Zacchei}, \& {Zonca}}]{Planck2020}
{Planck Collaboration}, {Aghanim}, N., {Akrami}, Y., {et~al.} 2020,
  \href{http://dx.doi.org/10.1051/0004-6361/201833910}{\JournalTitle{\aap},
  641, A6}

\bibitem[{{Rees} \& {Ostriker}(1977)}]{Rees1977}
{Rees}, M.~J., \& {Ostriker}, J.~P. 1977,
  \href{http://dx.doi.org/10.1093/mnras/179.4.541}{\JournalTitle{\mnras}, 179,
  541}

\bibitem[{Reines \& Volonteri(2015)}]{Reines2015RELATIONSUNIVERSE}
Reines, A.~E., \& Volonteri, M. 2015,
  \href{http://dx.doi.org/10.1088/0004-637X/813/2/82}{\JournalTitle{The
  Astrophysical Journal}, 813, 82}

\bibitem[{Rupke \& Veilleux(2013)}]{Rupke2013BREAKINGQSO}
Rupke, D. S.~N., \& Veilleux, S. 2013,
  \href{http://dx.doi.org/10.1088/2041-8205/775/1/L15}{\JournalTitle{The
  Astrophysical Journal}, 775, L15}

\bibitem[{{Sahu} {et~al.}(2019){Sahu}, {Graham}, \& {Davis}}]{Sahu2019a}
{Sahu}, N., {Graham}, A.~W., \& {Davis}, B.~L. 2019,
  \href{http://dx.doi.org/10.3847/1538-4357/ab0f32}{\JournalTitle{\apj}, 876,
  155}

\bibitem[{{Savorgnan} {et~al.}(2016){Savorgnan}, {Graham}, {Marconi}, \&
  {Sani}}]{Savorgnan2016}
{Savorgnan}, G. A.~D., {Graham}, A.~W., {Marconi}, A.~r., \& {Sani}, E. 2016,
  \href{http://dx.doi.org/10.3847/0004-637X/817/1/21}{\JournalTitle{\apj}, 817,
  21}

\bibitem[{{Schaye} {et~al.}(2015){Schaye}, {Crain}, {Bower}, {Furlong},
  {Schaller}, {Theuns}, {Dalla Vecchia}, {Frenk}, {McCarthy}, {Helly},
  {Jenkins}, {Rosas-Guevara}, {White}, {Baes}, {Booth}, {Camps}, {Navarro},
  {Qu}, {Rahmati}, {Sawala}, {Thomas}, \& {Trayford}}]{Schaye2015}
{Schaye}, J., {Crain}, R.~A., {Bower}, R.~G., {et~al.} 2015,
  \href{http://dx.doi.org/10.1093/mnras/stu2058}{\JournalTitle{\mnras}, 446,
  521}

\bibitem[{{Schutte} {et~al.}(2019){Schutte}, {Reines}, \&
  {Greene}}]{Schutte2019}
{Schutte}, Z., {Reines}, A.~E., \& {Greene}, J.~E. 2019,
  \href{http://dx.doi.org/10.3847/1538-4357/ab35dd}{\JournalTitle{\apj}, 887,
  245}

\bibitem[{{Shi} {et~al.}(2022){Shi}, {Kremer}, {Grudi{\'c}},
  {Gerling-Dunsmore}, \& {Hopkins}}]{Shi2022_hypereddington}
{Shi}, Y., {Kremer}, K., {Grudi{\'c}}, M.~Y., {Gerling-Dunsmore}, H.~J., \&
  {Hopkins}, P.~F. 2022, \JournalTitle{arXiv e-prints}, arXiv:2208.05025

\bibitem[{{Somerville} \& {Dav{\'e}}(2015)}]{SD2015_ARAA}
{Somerville}, R.~S., \& {Dav{\'e}}, R. 2015,
  \href{http://dx.doi.org/10.1146/annurev-astro-082812-140951}{\JournalTitle{\araa},
  53, 51}

\bibitem[{Springel {et~al.}(2005)Springel, Di~Matteo, \&
  Hernquist}]{Springel2005BlackGalaxies}
Springel, V., Di~Matteo, T., \& Hernquist, L. 2005,
  \href{http://dx.doi.org/10.1086/428772}{\JournalTitle{The Astrophysical
  Journal}, 620, L79}

\bibitem[{{Springel} \& {Hernquist}(2003)}]{SH03_multiphase}
{Springel}, V., \& {Hernquist}, L. 2003,
  \href{http://dx.doi.org/10.1046/j.1365-8711.2003.06206.x}{\JournalTitle{\mnras},
  339, 289}

\bibitem[{Stern {et~al.}(2020)Stern, Fielding, Faucher-Gigu{\`{e}}re, \&
  Quataert}]{Stern2020TheHaloes}
Stern, J., Fielding, D., Faucher-Gigu{\`{e}}re, C.-A., \& Quataert, E. 2020,
  \href{http://dx.doi.org/10.1093/MNRAS/STAA198}{\JournalTitle{Monthly Notices
  of the Royal Astronomical Society}, 492, 6042}

\bibitem[{Stern {et~al.}(2021{\natexlab{a}})Stern, Sternberg,
  Faucher-Gigu{\`{e}}re, Hafen, Fielding, Quataert, Wetzel,
  Angl{\'{e}}s-Alc{\'{a}}zar, El-Badry, Kere{\v{s}}, \&
  Hopkins}]{Stern2021NeutralRedshift}
Stern, J., Sternberg, A., Faucher-Gigu{\`{e}}re, C.~A., {et~al.}
  2021{\natexlab{a}},
  \href{http://dx.doi.org/10.1093/MNRAS/STAB2240}{\JournalTitle{Monthly Notices
  of the Royal Astronomical Society}, 507, 2869}

\bibitem[{Stern {et~al.}(2021{\natexlab{b}})Stern, Faucher-Gigu{\`{e}}re,
  Fielding, Quataert, Hafen, Gurvich, Ma, Byrne, El-Badry,
  Angl{\'{e}}s-Alc{\'{a}}zar, Chan, Feldmann, Kere{\v{s}}, Wetzel, Murray, \&
  Hopkins}]{Stern2021VirializationFeedback}
Stern, J., Faucher-Gigu{\`{e}}re, C.-A., Fielding, D., {et~al.}
  2021{\natexlab{b}}, \JournalTitle{The Astrophysical Journal}, 911, 88

\bibitem[{{Tacchella} {et~al.}(2015){Tacchella}, {Carollo}, {Renzini},
  {F{\"o}rster Schreiber}, {Lang}, {Wuyts}, {Cresci}, {Dekel}, {Genzel},
  {Lilly}, {Mancini}, {Newman}, {Onodera}, {Shapley}, {Tacconi}, {Woo}, \&
  {Zamorani}}]{Tacchella2015}
{Tacchella}, S., {Carollo}, C.~M., {Renzini}, A., {et~al.} 2015,
  \href{http://dx.doi.org/10.1126/science.1261094}{\JournalTitle{Science}, 348,
  314}

\bibitem[{Terrazas {et~al.}(2016)Terrazas, Bell, Henriques, White, Cattaneo, \&
  Woo}]{Terrazas2016QUIESCENCEGALAXIES}
Terrazas, B.~A., Bell, E.~F., Henriques, B. M.~B., {et~al.} 2016,
  \href{http://dx.doi.org/10.3847/2041-8205/830/1/l12}{\JournalTitle{The
  Astrophysical Journal}, 830, L12}

\bibitem[{{Tillman} {et~al.}(2021){Tillman}, {Wellons}, {Faucher-Gigu{\`e}re},
  {Kelley}, \& {Angl{\'e}s-Alc{\'a}zar}}]{Tillman_running_late}
{Tillman}, M.~T., {Wellons}, S., {Faucher-Gigu{\`e}re}, C.-A., {Kelley}, L.~Z.,
  \& {Angl{\'e}s-Alc{\'a}zar}, D. 2021, \JournalTitle{arXiv e-prints},
  arXiv:2109.14647

\bibitem[{{Toomre}(1977)}]{Toomre1977_mergers}
{Toomre}, A. 1977, in Evolution of Galaxies and Stellar Populations, ed. B.~M.
  {Tinsley} \& D.~C. {Larson}, Richard B.~Gehret, 401

\bibitem[{Torrey {et~al.}(2017)Torrey, Hopkins, Faucher-Gigu{\`{e}}re,
  Vogelsberger, Quataert, Kere{\v{s}}, \& Murray}]{Torrey2017AnNuclei}
Torrey, P., Hopkins, P.~F., Faucher-Gigu{\`{e}}re, C.~A., {et~al.} 2017,
  \href{http://dx.doi.org/10.1093/MNRAS/STX254}{\JournalTitle{Monthly Notices
  of the Royal Astronomical Society}, 467, 2301}

\bibitem[{{Tremaine} {et~al.}(2002){Tremaine}, {Gebhardt}, {Bender}, {Bower},
  {Dressler}, {Faber}, {Filippenko}, {Green}, {Grillmair}, {Ho}, {Kormendy},
  {Lauer}, {Magorrian}, {Pinkney}, \& {Richstone}}]{Tremaine2002}
{Tremaine}, S., {Gebhardt}, K., {Bender}, R., {et~al.} 2002,
  \href{http://dx.doi.org/10.1086/341002}{\JournalTitle{\apj}, 574, 740}

\bibitem[{Tremmel {et~al.}(2018)Tremmel, Governato, Volonteri, Pontzen, \&
  Quinn}]{Tremmel2018WanderingHalos}
Tremmel, M., Governato, F., Volonteri, M., Pontzen, A., \& Quinn, T.~R. 2018,
  \href{http://dx.doi.org/10.3847/2041-8213/AABC0A}{\JournalTitle{The
  Astrophysical Journal Letters}, 857, L22}

\bibitem[{{van de Voort} {et~al.}(2011){van de Voort}, {Schaye}, {Booth},
  {Haas}, \& {Dalla Vecchia}}]{vdV11}
{van de Voort}, F., {Schaye}, J., {Booth}, C.~M., {Haas}, M.~R., \& {Dalla
  Vecchia}, C. 2011,
  \href{http://dx.doi.org/10.1111/j.1365-2966.2011.18565.x}{\JournalTitle{\mnras},
  414, 2458}

\bibitem[{{van Dokkum} {et~al.}(2014){van Dokkum}, {Bezanson}, {van der Wel},
  {Nelson}, {Momcheva}, {Skelton}, {Whitaker}, {Brammer}, {Conroy},
  {F{\"o}rster Schreiber}, {Fumagalli}, {Kriek}, {Labb{\'e}}, {Leja},
  {Marchesini}, {Muzzin}, {Oesch}, \& {Wuyts}}]{vdK2014}
{van Dokkum}, P.~G., {Bezanson}, R., {van der Wel}, A., {et~al.} 2014,
  \href{http://dx.doi.org/10.1088/0004-637X/791/1/45}{\JournalTitle{\apj}, 791,
  45}

\bibitem[{Weinberger {et~al.}(2018)Weinberger, Springel, Pakmor, Nelson, Genel,
  Pillepich, Vogelsberger, Marinacci, Naiman, Torrey, \&
  Hernquist}]{Weinberger2018SupermassiveSimulation}
Weinberger, R., Springel, V., Pakmor, R., {et~al.} 2018,
  \href{http://dx.doi.org/10.1093/mnras/sty1733}{\JournalTitle{Monthly Notices
  of the Royal Astronomical Society}, 479, 4056}

\bibitem[{{Wellons} {et~al.}(2022){Wellons}, {Faucher-Gigu{\`e}re}, {Hopkins},
  {Quataert}, {Angl{\'e}s-Alc{\'a}zar}, {Feldmann}, {Hayward}, {Kere{\v{s}}},
  {Su}, \& {Wetzel}}]{Wellons2022_AGN}
{Wellons}, S., {Faucher-Gigu{\`e}re}, C.-A., {Hopkins}, P.~F., {et~al.} 2022,
  \JournalTitle{arXiv e-prints}, arXiv:2203.06201

\bibitem[{Wetzel {et~al.}(2016)Wetzel, Hopkins, Kim, Faucher-Gigu{\`{e}}re,
  Kere{\v{s}}, \& Quataert}]{Wetzel2016RECONCILINGGALAXY}
Wetzel, A.~R., Hopkins, P.~F., Kim, J.-h., {et~al.} 2016,
  \href{http://dx.doi.org/10.3847/2041-8205/827/2/L23}{\JournalTitle{The
  Astrophysical Journal Letters}, 827, L23}

\bibitem[{White \& Rees(1978)}]{White1978CoreClustering}
White, S. D.~M., \& Rees, M.~J. 1978,
  \href{http://dx.doi.org/10.1093/mnras/183.3.341}{\JournalTitle{Monthly
  Notices of the Royal Astronomical Society}, 183, 341}

\bibitem[{Xu \& Peng(2021)}]{Xu2021CriticalUniverseb}
Xu, B., \& Peng, Y. 2021,
  \href{http://dx.doi.org/10.3847/2041-8213/AC3A01}{\JournalTitle{The
  Astrophysical Journal Letters}, 923, L29}

\bibitem[{{Yu} {et~al.}(2021){Yu}, {Bullock}, {Klein}, {Stern}, {Wetzel}, {Ma},
  {Moreno}, {Hafen}, {Gurvich}, {Hopkins}, {Kere{\v{s}}},
  {Faucher-Gigu{\`e}re}, {Feldmann}, \& {Quataert}}]{Yu2021}
{Yu}, S., {Bullock}, J.~S., {Klein}, C., {et~al.} 2021,
  \href{http://dx.doi.org/10.1093/mnras/stab1339}{\JournalTitle{\mnras}, 505,
  889}

\bibitem[{{Yu} {et~al.}(2022){Yu}, {Bullock}, {Gurvich}, {Hafen}, {Stern},
  {Boylan-Kolchin}, {Faucher-Gigu{\`e}re}, {Wetzel}, {Hopkins}, \&
  {Moreno}}]{Yu2022_born_this_way}
{Yu}, S., {Bullock}, J.~S., {Gurvich}, A.~B., {et~al.} 2022,
  \JournalTitle{arXiv e-prints}, arXiv:2210.03845

\end{thebibliography}

\end{document}